\documentclass[11pt,x11names,a4paper]{article}
\pdfoutput=1

\usepackage[utf8]{inputenc}
\usepackage{lmodern}
\usepackage[T1]{fontenc} % Use 8-bit encoding that has 256 glyphs
\usepackage{microtype} % Slightly tweak font spacing for aesthetics

\usepackage[a4paper, left=25mm, right=25mm, top=30mm, bottom=25mm]{geometry} % Document margins
\usepackage{fancyhdr} % Headers and footers
\usepackage{indentfirst}
\usepackage{abstract}

\usepackage{titlesec} % Allows customization of titles
\titleformat{\section}[block]{\large\bfseries\centering}{\thesection}{1em}{} % Change the look of the section titles
\titleformat{\subsection}[block]{\bfseries}{\thesubsection}{1em}{} % Change the look of the section titles

\usepackage[dvipsnames]{xcolor}
\definecolor{celest}{RGB}{82, 24, 184}  
\definecolor{purplee}{RGB}{128, 56, 197}
\definecolor{spacec}{RGB}{28, 150, 150} 
\definecolor{applegreen}{rgb}{0.55, 0.71, 0.0}
\definecolor{caribbeangreen}{rgb}{0.0, 0.8, 0.6}
\definecolor{deepjunglegreen}{rgb}{0.0, 0.29, 0.29}
\definecolor{green(munsell)}{rgb}{0.0, 0.66, 0.47}
\definecolor{cerulean}{rgb}{0.0, 0.48, 0.65}
\definecolor{coolblack}{rgb}{0.0, 0.18, 0.39}
\definecolor{darkblue}{rgb}{0.0, 0.0, 0.55}
\definecolor{darkcerulean}{rgb}{0.03, 0.27, 0.49}
\definecolor{darkraspberry}{rgb}{0.53, 0.15, 0.34}
\definecolor{darkspringgreen}{rgb}{0.09, 0.45, 0.27}
\definecolor{tealblue}{rgb}{0.21, 0.46, 0.53}
\usepackage{cite}
\usepackage{hyperref}
\hypersetup{
	colorlinks=true,
	linkcolor=coolblack,
	citecolor=darkraspberry,
	urlcolor=darkspringgreen,
	linktoc=page
}
\newcommand{\fama}{family A }
\newcommand{\famb}{family B }
\newcommand{\al}{\alpha}
\newcommand{\g}{\gamma}
\newcommand{\de}{\delta}

\newcommand{\la}{\lambda}

\newcommand{\s}{\sigma}

\newcommand{\e}{\mathrm{e}}
\newcommand{\UU}{\mathrm{U}}

\newcommand{\R}{\mathbf{R}}

\newcommand{\be}{\begin{equation}}
\newcommand{\ee}{\end{equation}}

\newcommand{\f}[2]{\frac{#1}{#2}}

\usepackage{enumerate}
\usepackage{graphicx}
\usepackage[percent]{overpic}
\usepackage{tikz}
\usepackage{tensor}
\usepackage{subcaption}
\usepackage{booktabs}
\usepackage[many]{tcolorbox}
\usepackage{booktabs}
\tcbset{colback=blue!10!white, colframe=blue!50!black, 
	highlight math style= {enhanced, %<-- needed for the ’remember’ options
		colframe=red,colback=red!10!white,boxsep=0pt}
}

\usepackage{amsmath,amssymb,slashed,mathtools}
\usepackage{amsthm}
\numberwithin{equation}{section}
\usepackage{slashed}

\DeclareSymbolFontAlphabet{\amsmathbb}{AMSb}%

\usepackage{physics}
\usepackage{dsfont}

\title{\fontsize{20pt}{24pt}\selectfont\textbf{Universal holographic Wilson loops\\ in 3d SCFTs}\vspace{2mm}}

\author{
	\large{
		\href{mailto:ffg1m24@soton.ac.uk}{Fri{\dh}rik Freyr Gautason}$^1$ and  \href{mailto:alexianix@hi.is}{Alexia Nix}$^2$}\\[5mm]
	{}$^1${STAG Research Centre $\&$ Mathematical Sciences, University of Southampton,}\\ 
	{\normalsize University Road, Southampton SO17 1BJ, UK}\\[5mm]
	{}$^2${\normalsize University of Iceland, Science Institute}\\
	{\normalsize Dunhaga 3, 107 Reykjav{\'i}k, Iceland}\\[5mm]
}

\date{}

\begin{document}  
	%%%%%%%%%%%%%%%%%%%%%%%%%%%%%%
	{\hypersetup{urlcolor=black}\maketitle}
	\thispagestyle{empty}

	\begin{abstract}
			We study the vacuum expectation value of half-$\rm{BPS}$ Wilson loop operators in two families of superconformal $\mathcal{N}=2$ Chern-Simons-matter theories. The first family is dual to AdS$_{4}$ solutions in M-theory, while the second one has a dual description in massive type IIA string theory. Utilizing the properties of the underlying geometry, we provide a universal description for the semiclassical quantization of a probe M2-brane and fundamental string in the respective holographic dual geometries. As a result, we find the one-loop partition function of both the M2-brane and the string which leads to a prediction for the large $N$ behaviour of the Wilson loops in the dual SCFTs. For theories with M-theory duals, we conjecture the full perturbative completion as a ratio of Airy functions.
		
	\end{abstract}

	\newpage
	
	\setcounter{tocdepth}{2}
	\tableofcontents
	\newpage
	%----------------------------------------------------------------------------------------
	%	INTRODUCTION
	%----------------------------------------------------------------------------------------
	\section{Introduction}
\label{sec:introduction}
%%%%%%%%%%%%%%%%%%%%%%%%%%%%%%%%
Precision holography is a powerful tool that allows us to probe strongly coupled field theories by studying the weakly coupled dynamics of strings and branes in its gravitational dual. While the holographic correspondence finds applications in various dimensions, in this work we will focus on the AdS$_{4}$/CFT$_{3}$ correspondence, to gain insight into a broad class of three-dimensional superconformal Chern-Simons-matter theories.  By considering the large $N$ behaviour of the free energy on $S^{3}$ of these theories, we may classify these superconformal field theories (SCFTs) into two families. Family A contains the $\mathcal{N}\geq2$ field theories whose degrees of freedom scale as $N^{3/2}$ such as the celebrated ABJM theory \cite{Aharony:2008ug}, while family B includes the $\mathcal{N}=2$ SCFTs whose free energy scales as $N^{5/3}$ in the large $N$ limit, see e.g. \cite{Gaiotto:2009mv}.  On the gravity side, this distinction manifests as family A field theories admitting a dual description in M-theory on AdS$_{4}\times$SE$_{7}$ where the internal space is a seven-dimensional Sasaki-Einstein manifold. On the other hand, the field theories of family B we consider are dual to massive type IIA string theory on a warped product of AdS$_{4}$ and $X_{6}\simeq\mathcal{S}(\text{SE}_{5})$, which is topologically a sine-cone over a five dimensional Sasaki-Einstein manifold \cite{Guarino:2015jca,Fluder:2015eoa}.\footnote{One can reduce eleven-dimensional M-theory to type IIA string theory by performing a double dimensional reduction \cite{Duff:1987bx}, while it was shown in \cite{Aharony:2010af} that massive type IIA cannot be uplifted to M-theory.} 

One way of accessing the strongly coupled dynamics of these field theories is to use supersymmetric localization \cite{Pestun:2007rz,Kapustin:2009kz,Hama:2010av}. For some observables, this reduces the problem of evaluating path integrals to the study of finite-dimensional matrix integrals. These may be simplified even further by focusing on the planar limit. 
As a result, one can extract the leading-order large $N$ behaviour of various observables, such as the sphere free energy and the vacuum expectation value (vev) of BPS Wilson loop (WL) operators on $S^{3}$. Both of these observables can also be computed in the same leading-order large $N$ limit using the dual description. Comparing the two results serves as a quantitative test of the AdS$_{4}$/CFT$_{3}$ correspondence, and has been the subject of many recent works. 
Studying subleading corrections to the large $N$ behaviour of these (or other holographic) observables either on the field theory or the gravitational side is a non-trivial task and often requires a case-specific treatment. 
In this work we focus on the 1/2-BPS WL from the gravitational perspective and show that there is no need for a case-by-case analysis; a universal treatment of the gravity duals of the three-dimensional SCFTs in \fama and \famb follows from the semiclassical quantization of an M2-brane and of a fundamental string, respectively.

It has been known for a long time that the gravitational dual to a Wilson loop is given by a fundamental string that attaches to the boundary as dictated by the field theory operator \cite{Maldacena:1998im}. The string partition function with these boundary conditions should match the vacuum expectation value of the dual operator. In practice, in the classical $L/\ell_s\gg1$ limit ($L$ is the AdS length scale), the partition function is dominated by the  classical string action and the one-loop quantization of the string gives rise to the leading $L/\ell_s$ correction \cite{Drukker:2000ep}. In this paper, we will use this approach to access the leading correction to the WL vacuum expectation value for theories in family B.

In contradistinction to the various holographic results of family A that we review below, quantitative tests of the AdS$_{4}$/CFT$_{3}$ correspondence for field theories belonging to family B are more limited. The first explicit example of such a test was provided by \cite{Guarino:2015jca}, where the free energy on $S^{3}$ of the $\mathcal{N}=2$ Chern-Simons-matter theory with gauge group $\text{SU}(N)_{k}$, in the large $N$ limit was found to match the on-shell classical supergravity action. This result was subsequently extended to a wide class of field theories and corresponding supergravity solutions in \cite{Fluder:2015eoa}. It was shown that both the leading order free energy and 1/2-BPS WL can be obtained from a matrix model analysis and the results match the respective supergravity calculation.  However, subleading corrections to these results have not yet been obtained in either gravity or field theory. 
In this work, we perform the one-loop quantization of the string identified in \cite{Fluder:2015eoa} and obtain the following prediction for the WL of family B
\begin{gather}\label{eq: fam B main result}
	\langle W\rangle_{\rm{B}}\approx  \Gamma\qty(\frac{2}{3})^{3}\frac{n^{2/3}N^{1/3}}{2^{4/3}3^{2/3}\pi\,\text{vol}(\text{SE}_{5})^{1/3}}\exp\qty(\pi^{2}\frac{4^{1/3}N^{1/3}}{3^{1/6}n^{1/3}\,\text{vol}(\text{SE}_{5})^{1/3}})\,,
\end{gather}
where $n$ is the sum over all Chern-Simons levels of the dual field theory. This is a universal answer that holds for all SE$_{5}$ manifolds and all corresponding field theory duals.

The string dual to Wilson loops can be uplifted to an M2-brane in M-theory for holographic backgrounds in \emph{massless} type IIA supergravity. In recent works, it has been demonstrated that the one-loop quantization of M2-branes is remarkably powerful. First, it is relatively straightforward to regularize the divergent partition function, and more importantly in the case where it has been tested, namely AdS$_4\times S^7/{\mathbf{Z}_k}$ dual to ABJM theory, it matches field theory results \cite{Giombi:2023vzu}. 
Motivated by this success, using similar methods we study the AdS$_{4}\times$SE$_{7}$ holographic duals to SCFTs in family A. We build on the work of \cite{Farquet:2013cwa} where the supersymmetric M2-brane configuration was identified and the role of the M-theory circle $S^{1}_{M}\subset \rm{SE}_{7}$ was highlighted. We show that the appropriate calibration of this M-theory circle plays a vital role in our one-loop analysis. More concretely, using this together with the geometric properties of Sasaki-Einstein manifolds we find a universal expression for the one-loop M2-brane partition function which depends only on the radius of the M-theory circle, which we denote by $c$, and three charges $q_{l}$, with $l=1,2,3$.
Importantly, $c$ and $q_{l}$ can be evaluated by using the explicit internal SE$_{7}$ metric of the supergravity solution and hence one can utilize our formula (which is given in (\ref{eq:gamma with qs} - \ref{M2 part funct})) for an abundance of Chern-Simons-matter theories that enjoy a known gravity dual. 
Inspired by the recent discussion regarding the choice of ensembles in AdS$_4$ holography \cite{Gautason:2025plx}, we speculate that the M2-brane partition function is one-loop exact albeit computed in the grand canonical ensemble. Upon mapping back to the canonical ensemble, we conjecture a formula that is perturbatively exact in $N$. This expression generalizes a similar expression that is available for the ABJM theory  \cite{Klemm:2012ii} which features a ratio of Airy functions. 
Given the very scarce results for the vacuum expectation value of the WLs for these SCFTs, our results serve as predictions which can in principle be checked  by a direct analysis in the matrix model on a case-by-case basis.

The remainder of this paper is structured as follows. In Section \ref{sec:2SCFTfams}, we review the matrix model of the two families of SCFTs we study and define the $1/2$-BPS Wilson loop. In Section \ref{sec:famA gravity duals} we provide a universal expression for the one-loop M2-brane partition function, which yields a prediction for the Wilson loop of \fama at subleading order. In Section \ref{sec:examples famA} we apply our results to some examples of \fama Chern-Simons-matter theories. We continue in Section \ref{sec:famB gravity duals} with the calculation of the one-loop string partition function of the gravity duals to family B, providing a prediction for the sub-leading correction to the $1/2$-BPS Wilson loop of this class of SCFTs. We conclude in Section \ref{sec:summary} with a summary of our results and discuss future directions. In Appendix \ref{app:SE} we collect some useful geometric properties of Sasaki-Einstein manifolds and in Appendix \ref{app:HK} we review the Heat kernel method. Finally, in Appendix \ref{app:spectrum fam B} we provide additional details on the derivation of the one-loop string partition function that we obtained for family B.

%----------------------------------------------------------------------------------------
%	Review of SCFTs/Matrix model 
%----------------------------------------------------------------------------------------
\section{Two families of Chern-Simons-matter theories}
\label{sec:2SCFTfams}
%%%%%%%%%%%%%%%%%%%%%%%%%%%%%%%%
In this section we briefly review the two classes of three-dimensional superconformal Chern-Simons-matter theories on $S^{3}$ which we study in this paper. We focus on vacuum expectation values of supersymmetric Wilson loop operators that can be computed using supersymmetric localization \cite{Pestun:2007rz, Kapustin:2009kz, Jafferis:2010un, Hama:2010av}. We consider superconformal gauge theories with gauge group $G$ that is a product of $\mathcal{G}$ unitary groups $G=\prod_{i=1}^{\mathcal{G}}\UU(N)_{k_{i}}$. Each $\UU(N)$ factor is associated with a corresponding Chern-Simons (CS) level $k_{i} \in \mathbf{Z}$. It is convenient to denote the sum of all CS levels by $n=\sum_{i=1}^{\mathcal{G}}k_{i}$. In addition to the vector multiplets that contain the gauge bosons, the theories are also equipped with chiral multiplets that transform in the adjoint representation of one of the gauge nodes or in a bifundamental representation of a pair of nodes. The gauge group, CS levels and matter content can be neatly packaged in a quiver diagram which consists of nodes denoting the gauge groups connected by directed lines denoting the bifundamental (or adjoint when the line begins and ends on the same node) matter multiplets. 

We now separate the SCFTs into two families. Family A consists of theories  for which the sum of CS level vanishes; $n=0$.	These theories enjoy a holographic dual description in M-theory as the near-horizon geometry of M2-branes probing CY$_4$ singularities. The prime example is the ABJM theory which features two gauge nodes with opposite CS levels and bifundamental matter fields \cite{Aharony:2008ug}. 
We will denote theories that have $n\ne 0$ as being in family B. These theories find a holographic description as the near-horizon geometry of D2-branes in massive type IIA string theory \cite{Gaiotto:2009mv}. For this latter class, following \cite{Fluder:2015eoa}, we will make the additional assumption that all Chern-Simons levels are equal to $k$ and thus $n= {\cal G}k$.

\subsection{The matrix model}
With this classification in place, we now turn to localization and the associated matrix model formulation. To this end we analytically continue to Euclidean signature and perform a conformal transformation to $S^3$.  
Next, we use the tools of supersymmetric localization to reduce the infinite-dimensional path integral to a finite-dimensional matrix integral. In simplified terms, this is done by choosing a suitable supercharge $Q$ and studying the equivariant cohomology of the chosen charge. In particular $Q$-closed observables can be computed by deforming the action by a $Q$-exact functional multiplied by $1/\hbar$. Since the deformation is $Q$-exact and the action is $Q$-closed, the observable is independent of $\hbar$. This means that we can compute BPS observables in the cohomology of $Q$ using the saddle point approximation $\hbar\to0$ which is one-loop exact. For the superconformal field theories on $S^{3}$ discussed above, the chiral multiplets vanish on the saddle point configuration (i.e. the localization locus) and we are left with the vector multiplets \cite{Kapustin:2009kz,Jafferis:2010un,Hama:2010av}. The $\mathcal{N}=2$ superconformal vector multiplet consists of the gauge field $A_{\mu}$, two auxiliary fields $\s$ and $D$ which are real scalars and the two-component complex spinors $\lambda$. On the localization locus, the gauge boson and fermion both vanish whereas the auxiliary scalars $\s$ and $D$ are related to each other and are constant.
Hence, the path integral reduces to an integral over ${\cal G}$ unitary $N\times N$ matrices $\s$, one for each gauge node. These can be diagonalized and parametrized in terms of their eigenvalues $\lambda_{I}/(2\pi)$, with $I=1,\dots,N$. This introduces the well-known Vandermonde determinant in the measure factor, which however cancels with a factor in the one-loop determinant that arises from the quadratic fluctuations of the fields around the localization locus. The resulting partition function on $S^{3}$ around this configuration then reads \cite{Kapustin:2009kz, Hama:2010av} 
\begin{gather}\label{eq:part func}
Z_{S^{3}}=\frac{1}{(N!)^{\mathcal{G}}}\int\prod_{j=1}^{\mathcal{G}}\bigg(\prod_{I=1}^{N}\frac{\dd{\lambda_{I}^{j}}}{2\pi}\exp(i\lambda_{I}^{j\,2}\frac{k_{j}}{4\pi})\prod_{I\neq J} 2\sinh\frac{\lambda_{IJ}^{j}}{2}\bigg)\prod_{m} \prod_{I,J=1}^{N}s(1+i\lambda_{IJ}^{m}-\Delta_{m})\,,
\end{gather}
where we defined $\lambda_{IJ}^{j}\equiv\lambda_{I}^{j}-\lambda_{J}^{j}$ and 
\begin{gather}
s(z)= \e^{-z\log(1-\e^{2\pi iz})-\frac{i\pi}{12}+\frac{i}{2\pi}\text{Li}_{2}(\e^{2\pi i z})+\frac{i\pi}{2}z^{2}}\,.
\end{gather} 
The exponential in \eqref{eq:part func} encodes the classical contribution from the Chern-Simons action evaluated on the localization locus. The second factor comes from the one-loop determinant that results from the quadratic fluctuations of the fields around the saddle. The last term of \eqref{eq:part func} captures the contribution from the chiral multiplets labelled by $m$ with $R$-charge $\Delta_{m}$. 
The matter sector has no classical contribution and there is no interaction between chiral and gauge fields at quadratic order. This leads to the factorization of the partition function into separate contributions coming from each sector. We note that this last term contains the one-loop contribution  from the chiral multiplet that transforms in the adjoint representation of $U(N)_{k_{m}}$, as well as the bifundamental chiral multiplet that transforms in the $(N,\bar{N})$ representation of $U(N)_{k_{n}}\times U(N)_{k_{l}}$ with corresponding $R$-charge $\Delta_{nl}$. More concretely, the fields transforming in the adjoint representation of $U(N)_{k_{m}}$ contribute as
\begin{gather}
\prod_{I,J=1}^{N}s(1+i\lambda_{IJ}^{m}-\Delta_{m})\,,
\end{gather}
while the chiral fields transforming in the bifundamental representation of $U(N)_{k_{n}}\times U(N)_{k_{l}}$ contribute as
\begin{gather}
\prod_{I,J=1}^{N}s(1+i(\lambda_{I}^{n}-\lambda_{J}^{l})-\Delta_{nl})\,,
\end{gather}
in \eqref{eq:part func}. 

In addition to the sphere partition function itself, supersymmetric localization allows for the computation of the expectation values of certain supersymmetric Wilson loop operators. These operators can be expressed as 
	\begin{gather}\label{eq:WL op}
		W_{\mathcal{R}}=\Tr_{\mathcal{R}}\Biggl\{\mathcal{P}\exp\Big(\oint_{\gamma}(iA_{\mu}\dot{x}^{\mu}+\s\abs{\dot{x}})\dd{s}\Big)\Biggr\}\,,
	\end{gather}
	where $\mathcal{R}$ is a representation of $G$, $\sigma$ is as before a scalar in the vector multiplet, and $\mathcal{P}$ is the necessary path-ordering around the loop $\gamma\subset S^{3}$ which is parametrized by $x^{\mu}(s)$. In order to preserve the localizing supercharge,  $\gamma$ is identified with the great circle $S^{1}\subset S^{3}$. This choice in fact renders the operator $1/2$-BPS.

\subsection{Large \texorpdfstring{$N$}{N} limit}
Having reviewed the matrix model, let us now turn to its large $N$ limit.\footnote{Please note that in this limit other parameters such as the CS levels, and number of fundamental chiral multiplets are kept fixed.} This was originally studied for the ABJM theory in \cite{Drukker:2010nc} and more generally for theories in family A, starting with \cite{Herzog:2010hf}. For the theories in \famb with equal CS levels, the large $N$ analysis was performed in \cite{Fluder:2015eoa}. 
The techniques used are quite similar for \fama and \famb although the results are very different. The first step is to propose an ansatz for the $N$-scaling of the eigenvalues $\lambda_{I}^{j}$ at large $N$. In general the saddle point equations of the matrix model can be complex and thus the large $N$ behaviour of the eigenvalues can be expressed in terms of its real and imaginary parts
\begin{gather}\label{eq:eigens large N}
\lambda_{I}^{j}=N^{\al}x_{I}+i N^{\beta}y_{I}^{j}\,,
\end{gather}  
where $\alpha,\beta\in\mathbf{R}$ and $x_{I}$, $y_{I}^{j}$ are assumed not to scale with $N$. Notice that the real parts of the eigenvalues are independent of the gauge node in question. This follows from the large $N$ analysis of the models \cite{Herzog:2010hf,Fluder:2015eoa}. In fact, for \famb one can also show that $y_{I}^{j}=y_{I}$. Similarly one can determine the exponents $\alpha$ and $\beta$ by a straight-forward scaling argument. For theories in \fama it was found in \cite{Herzog:2010hf,Martelli:2011qj} that $\alpha=1/2$ and $\beta=0$, while \famb has $\al=\beta=1/3$ \cite{Fluder:2015eoa}.  Now, one can introduce an eigenvalue density 
\begin{gather}\label{eq:density eigens}
\rho(x)=\frac{1}{N}\sum_{I=1}^{N}\delta(x-x_{I})\,,
\end{gather}
which in the large $N$ limit is approximately given by a continuous function with compact support. By definition the density is normalized such that
\begin{gather}\label{eq:int density eigens}
\int_{-x_{*}}^{x_{*}}\rho(x)\dd{x}=1\,,
\end{gather}  
where we denoted the endpoints of the eigenvalue distribution by $\pm x_{*}$. It is also useful to parametrize the imaginary part of the eigenvalue in terms of its real part, i.e. $y^{j} \equiv y^j(x)$. Similar to the density, this is an approximately continuous function in the large $N$ limit. Finally, in the same limit, the partition function is dominated by its saddle point and so we extremize the free energy $F_{S^3} = -\log Z_{S^3}$ with respect to the eigenvalue density $\rho(x)$ and the functions $y^j(x)$ subject to the normalization condition \eqref{eq:int density eigens}. This will yield a non-trivial saddle only if all contributing terms in $F_{S^3}$ scale in the same way in the large $N$ limit which is how the scaling exponents $\alpha$ and $\beta$ we quoted above were determined. The result for the eigenvalue density itself depends on the theory in question but in general for \fama the density is found to be a piecewise linear function \cite{Herzog:2010hf} whereas for \famb it is a quadratic \cite{Fluder:2015eoa}.

Once these expressions have been determined, they can be inserted back into the formula for the free energy to give its large $N$ approximation. In line with the expectation from holography, it was shown that the free energy can be expressed in terms of the geometric data of the dual supergravity background (if available). More precisely, for generic holographic theories of family A, the holographic dual background takes the form AdS$_{4}\times$SE$_{7}$ where SE$_7$ denotes a Sasaki-Einstein space. The free energy can be computed in holography by evaluating the regularized on-shell action on these backgrounds. The result is the simple expression \cite{Herzog:2010hf} 
\begin{gather}\label{eq:free energy LO fama}
F_{S^{3}}^{A}= \sqrt{\frac{2\pi^{6}}{27\,\text{vol(SE$_{7}$)}}}N^{3/2}\,.
\end{gather}
This result was reproduced from the matrix model in the large $N$ limit as described above by mapping the  volume of the Sasaki-Einstein space to a function of the R-charges of the dual field theory \cite{Herzog:2010hf,Martelli:2011qj}.

The story is similar for family B. However, here the holographic background is of the form AdS$_{4}\times{\cal S}(\text{SE}_{5})$ where ${\cal S}(\text{SE}_{5})$ denotes the so-called sine-cone over a five-dimensional Sasaki-Einstein space (we refer to Section \ref{sec:famB gravity duals} for more details) \cite{Guarino:2015jca,  Fluder:2015eoa}. Carrying out the large $N$ limit of the matrix model to compute the free energy, one recovers the holographic result \cite{Guarino:2015jca, Jafferis:2011zi, Fluder:2015eoa} 
	\begin{gather}\label{eq:free energy LO famb}
		 F_{S^{3}}^{B}=\frac{3^{1/6}2^{1/3}\pi^{3}n^{1/3}}{5\,\text{vol(SE$_{5}$)}^{2/3}}N^{5/3}\,.
	\end{gather}

Just like the free energy, the vacuum expectation value of the Wilson loop can also be computed in a straight-forward manner once the large $N$ density of eigenvalues has been determined. Its evaluation boils down to an insertion of $\sum_{j=1}^{\mathcal{G}}\sum_{I=1}^{N}\e^{\lambda_{I}^{j}}$ in the original matrix model which in the large $N$ limit, reduces to
	\begin{gather}
		\langle W\rangle_{\text{SCFT}}=N\sum_{j=1}^{\mathcal{G}}\int_{-x_{*}}^{x_{*}}\dd{x}\rho(x)\e^{N^{\al}x+i N^{\beta}y(x)^{j}}\,.
	\end{gather}
	This integral is dominated by its contribution from the positive endpoint, which allows us to write
	\begin{gather}\label{eq:WL large N MM}
			\Re\log\langle W\rangle_{\text{SCFT}}=x_{*}N^\alpha\,,
	\end{gather}
	to leading order in large $N$.
	Using the result for the eigenvalue density and its endpoints, the leading order contribution to the Wilson loop for each family is \cite{Farquet:2013cwa,Fluder:2015eoa}
	\begin{gather}
		\log\langle W\rangle_{A}= 2\pi^{3}c\sqrt{\frac{N}{6\,\text{vol}(\text{SE}_{7})}}\,, \label{eq:WL LO fama} \\ \Re\log\langle W\rangle_{B}= \pi^{2}\qty(\frac{4N}{\sqrt{3}n\,\text{vol}(\text{SE}_{5})})^{1/3}\,,\label{eq:WL LO famb}
	\end{gather}
where we once more expressed the field theory data in terms of the geometric data of the dual supergravity solution. As will be explained in Section \ref{sec:famA gravity duals}, the parameter $c$ in \eqref{eq:WL LO fama} is the appropriately normalized radius of the M-theory circle $S^{1}_{M}$ which the M2-brane wraps in the dual eleven-dimensional geometry.

In this paper our focus is on the next-to-leading order contribution to the Wilson loop vev. This is notoriously complicated to compute in the matrix model and for most of the examples is not known. The leading order density does not suffice when computing this subleading correction to the Wilson loop vev. Naively one might have assumed that it is relatively straight-forward to compute the required correction, however, it turns out that there is an important correction that arises at the edge of the eigenvalue distribution that has to be carefully determined (see e.g. \cite{Chen-Lin:2014dvz}). We will not pursue this direction in this paper but instead take an alternative approach. We will use holography to directly evaluate the Wilson loop vev to next-to-leading order which will serve as a prediction for future studies of the matrix model.

\subsection{The Airy conjecture}
In \cite{Marino:2011eh} it was shown that for a large class of ${\cal N}=3$ Chern-Simons matter SCFTs whose degrees of freedom scale as $N^{3/2}$,  the perturbative part of the sphere partition function sums to an Airy function. Subsequently, in \cite{Marino:2011eh}, the same authors found supporting evidence for this conjecture for a specific class of $\mathcal{N}=2$ SCFTs.\footnote{In this work we focus on the superconformal and round sphere scenario. We refer the interested reader to \cite{Bobev:2025ltz} and references therein for a discussion of a more general Airy conjecture and evidences for it.} 
Concretely, for our purpose, the Airy conjecture states that (the perturbative part of) the  partition function on $S^{3}$ for a large class of three-dimensional  $\mathcal{N}\geq2$ SCFTs is given by the following expression
\begin{gather}\label{airy conj}
Z^{\text{p}}_{\text{SCFT}}(N)=\mathcal{C}^{-1/3}\e^{\mathcal{A}}\text{Ai}(\mathcal{C}^{-1/3}(N-\mathcal{B}))\,.
\end{gather}
Here the $\mathcal{A}$, $\mathcal{B}$ and $\mathcal{C}$ parameters are independent of the rank of the gauge group $N$ and are only a function of the Chern-Simons level $k$, the number of fundamental and anti-fundamental chiral multiplets, as well as the $R$-charges of the chiral multiplets. 
Combining \eqref{eq:free energy LO fama} with the large $N$ limit of \eqref{airy conj} yields the following expression for the $\mathcal{C}$ parameter in terms of the volume of the SE$_{7}$ manifold featured in the holographic dual 
\begin{gather}\label{Ck in terms of SE}
\mathcal{C}=\frac{6\text{vol}(\text{SE}_{7})}{\pi^{6}}\,,
\end{gather}
which we highlight is independent of the quantity $c$ encountered in the Wilson loop expectation value. 
The parameters ${\cal A}$ and ${\cal B}$ are much more difficult to determine and are only known in rare cases.
For reference, we list the known parameters in the argument of the Airy function for the SCFTs we study in Section \ref{sec:examples famA} in Table \ref{tab:airy param} below. Entries for the $\mathcal{B}$ parameter that have not yet been determined, are indicated in the table with a dash.

\begin{table}[h!]
	\centering
	\begin{tabular}{lccc}
		\toprule
		SCFT & $\mathcal{B}$ & $\mathcal{C}$ & Reference \\
		\midrule
		ABJM &  $\frac{1}{3k}+\frac{k}{24}$  & $\frac{2}{\pi^{2}k}$ &\cite{Fuji:2011km,Marino:2011eh}\\[2pt]
		ADHM &  $\frac{1}{2N_{f}}-\frac{N_{f}}{8}$  & $\frac{2}{\pi^{2}N_{f}}$ & \cite{Jafferis:2011zi,Mezei:2013gqa, Grassi:2014vwa}\\[2pt]
		$Q^{1,1,1}/\mathbf{Z}_{N_{f}}$  & $\frac{1}{4N_{f}}-\frac{N_{f}}{12}$  & $\frac{3}{4\pi^{2}N_{f}}$ &\cite{Cheon:2011vi,Jafferis:2011zi,Geukens:2024zmt,Bobev:2025ltz}\\[2pt]
		$Q^{1,1,1}/\mathbf{Z}_{k}$  & -  & $\frac{3}{4\pi^{2}k}$ & - \\[2pt]
		$Q^{2,2,2}/\mathbf{Z}_{k}$  & -  & $\frac{3}{8\pi^{2}k}$ & \cite{Amariti:2011uw} \\[2pt]
		$V_{5,2}/\mathbf{Z}_{N_{f}}$ & $\frac{5}{16N_{f}}+\frac{N_{f}}{48}$  & $\frac{81}{64\pi^{2}N_{f}}$ &\cite{Cheon:2011vi,Jafferis:2011zi,Geukens:2024zmt,Bobev:2025ltz}\\[2pt]
		$V_{5,2}/\mathbf{Z}_{k}$ & -  & $\frac{81}{64\pi^{2}k}$ &\cite{Cheon:2011vi, Martelli:2011qj}\\[2pt]
		$M^{3,2}/\mathbf{Z}_{k}$ & - & $\frac{27}{64\pi^{2}k}$ & \cite{Amariti:2011uw,Gang:2011jj} \\
		\bottomrule
	\end{tabular}
	\caption{Summary of known Airy function parameters for the examples studied in Section \ref{sec:examples famA}. The reference refers to the matrix model calculation of the $\mathcal{B},\mathcal{C}$ parameters. We note that the $\mathcal{C}$ parameter of the $Q^{1,1,1}/\mathbf{Z}_{k}$ theory has not been derived from the matrix model.}
	\label{tab:airy param}
\end{table}
	
In line with the universal form of the $S^{3}$ partition function \eqref{airy conj}, in this paper we will provide evidence for a universal expression of the vacuum expectation value of a $1/2$-BPS Wilson loop for the \fama SCFTs (c.f. \eqref{WL prop}). We achieve this using holography, by studying the partition function of a supersymmetric M2-brane wrapping an $S^{1}_{M}$ circle in the eleven-dimensional dual gravity background at one-loop order. This will give rise to a prediction for the Wilson loop vev beyond leading order and will inspire us to conjecture its perturbatively exact expression in Section \ref{sec:WL prop}. We now turn to this analysis.

\section{Holographic Wilson loop for \fama} 
\label{sec:famA gravity duals}
In this section we extract the prefactor to the exponential behaviour of the Wilson loop operator in the \fama  superconformal field theories we reviewed in Section \ref{sec:2SCFTfams}. We achieve this through holography, where the Wilson loop vev on $S^3$ can be computed as the partition function of an M2-brane in AdS$_4\times \text{SE}_{7}$ which extends to the boundary of AdS$_4$ and wraps the so-called M-theory circle which we discuss in more detail. We fix the boundary conditions of the M2-brane by matching its position on the $S^3$ to the position of the dual Wilson loop whereas its location in the extra $\text{SE}_{7}$ direction is determined by the symmetry preserved by the loop. The starting point is to identify a suitable classical configuration for the M2-brane with these boundary conditions. Fortunately, this task has already been solved in \cite{Farquet:2013cwa} and we review their results below. The remarkable feature of their computation is that they did not specify an explicit Sasaki-Einstein manifold $\text{SE}_{7}$, but instead find a general result that is applicable to \emph{any} supersymmetric AdS$_4\times \text{SE}_{7}$.\footnote{In fact, \cite{Farquet:2013cwa} also study warped AdS$_{4}$ solutions in M-theory. In this work we only consider the specific case where the background is given by an AdS$_{4}\times \text{SE}_{7}$ solution.} On the field theory side this implies that the leading contribution to the Wilson loop in the large $N$ limit can be given in terms of simple geometric data of the dual M-theory solution.

After reviewing the classical M2-brane solution, our next task is to compute the one-loop quantization of the M2-brane and thereby determine the coefficient of the exponential in the dual Wilson loop vev. Again, we observe that the result is completely universal and we are not forced to pick a particular Sasaki-Einstein manifold when carrying out our computation.

\subsection{The classical M2-brane}
\label{sec:QuantumM2}
The eleven-dimensional supergravity backgrounds dual to the SCFTs of family A we consider are given by 
\begin{eqnarray}\label{11d metric}
	\dd{s}^{2}_{11}&=&\frac{R^{2}}{4}\dd{s}_{\text{AdS}_{4}}^{2}+R^{2}\dd{s}_{\text{SE}_{7}}^{2}\,,\\
	\dd{s}_{\text{SE}_{7}}^{2}&=&\eta^{2}+\dd{s}_{\text{KE}_{6}}^{2}\,,
	\label{SE metric}\\
	G_{4}&=&\frac{3}{8}R^{3}\text{vol}_{\text{AdS}_{4}}\,,
\end{eqnarray}
where $\dd{s}_{\text{SE}_{7}}^2$ is the metric on `unit-radius' Sasaki-Einstein manifold for which  \(R_{\mu\nu}^{(\text{SE}_{7})}=6g_{\mu\nu}^{(\text{SE}_{7})}\), and the Kähler two-form $\omega$ associated to the transverse six-dimensional Kähler-Einstein manifold is given by \(\dd{\eta}=2\omega\). Notice that the overall scale of the geometry is set by the radius $R$. In particular the length scale of AdS is $L=R/2$. The supergravity approximation is valid whenever this radius is large compared to the eleven-dimensional Planck length $\ell_p$.

As mentioned above, the vacuum expectation value of a BPS Wilson loop is captured by the partition function of probe M2-brane in this background. 
In the limit where $R/\ell_p$ is large, we can approximate the partition function by its contribution from the dominant saddle. The latter is computed by finding a classical configuration of the brane in the geometry (consistent with its boundary conditions) and evaluating
\begin{gather}
	\langle W \rangle=Z_{\text{M2}} \approx \e^{-S_\text{cl}}Z_{1-\text{loop}}\,,
\end{gather}
where $S_\text{cl}$ is the (regularized) on-shell action of the classical M2-brane configuration and $Z_{1-\text{loop}}$ is its one-loop partition function. Of course this expression can be supplemented by higher loop corrections, but these are usually out of reach and will not be considered in this paper.

Let us now review the results of \cite{Farquet:2013cwa} which identified the classical configuration of the M2-brane and computed its classical action. 
The M2 supermembrane action was first constructed in \cite{Bergshoeff:1987cm}, but for now we focus only on its bosonic terms which can be grouped into the Polyakov-type action
\begin{gather}\label{eq:M2brane bos action}
	S_{\text{M2}}^{\text{bos}}=\frac{1}{2(2\pi)^{2}\ell_{p}^{3}}\int(\hat\gamma^ {ij}\hat G_{ij}-1)\text{vol}_{3}-\frac{i}{(2\pi)^{2}\ell_{p}^{3}}\int {A}_{3}\,,	
\end{gather}
where $\hat\gamma_{ij}$ is the worldvolume metric of the M2-brane (and thus we reserve the use of the Latin indices $i,j=1,2,3$ for the M2-brane worldvolume), vol$_{3}$ denotes its volume form and the target space metric is given by $\hat G_{MN}$. Note that we use hats to distinguish the eleven-dimensional objects from their ten-dimensional counterparts which we encounter later in this paper. The pull-back of the target space metric is denoted by $\hat G_{ij}= \hat G_{MN}\partial_iX^M\partial_j X^N$ where $X^M$ are the embedding functions of the brane into the target space. The last term in the bosonic action captures the coupling of the M2-brane to the target space three-form field $A_3$, and here by a slight abuse of notion we use the same symbol to denote the target space field as well as its pull-back to the worldvolume. 
We work in static gauge and thus identify three of the target space coordinates with the worldvolume coordinates. At the classical level, the pull-back of the target space metric to the worldvolume then yields $\hat G_{ij}^\text{cl}=\hat\gamma_{ij}$.

The classical configuration of the M2-brane dual to the supersymmetric Wilson loop is simply a direct product $\text{AdS}_{2}\times S^{1}_{M}$ where  $\text{AdS}_{2}$ is a minimal surface inside AdS$_4$ and is familiar from the corresponding minimal string configuration in type IIA string theory. Compared to the fundamental string, the M2-brane also wraps some a priori unspecified M-theory circle $S^{1}_{M}\subset\text{SE}_{7}$. Let $\zeta_M$ denote the Killing vector that generates the M-theory circle. It is important that this is not the so-called Reeb vector $\zeta_{R}$ that generates the $\UU(1)_R$ dual to the field theory R-symmetry.\footnote{Recall that the Reeb vector $\zeta_{R}$ satisfies $\zeta_{R}\lrcorner\eta=1$ and $\zeta_{R}\lrcorner\dd\eta=0$.} Having chosen the M-theory circle $\UU(1)_M$, in order to fully specify the classical M2-brane configuration, we must specify its location in the transverse six-dimensional manifold $\text{SE}_{7}/\UU(1)_M$. As discussed in \cite{Farquet:2013cwa}, supersymmetry dictates that the M2-brane sits at points $p\in \text{SE}_{7}/\UU(1)_M$ where $\zeta_M$ is proportional to the Reeb vector $\zeta_{R}$. This can also be phrased differently. Following \cite{Farquet:2013cwa} we define a Hamiltonian function $h_{M}:\text{SE}_{7}\to \R$ as 
\begin{gather}\label{eq:ham funct}
	h_{M}\equiv \zeta_{M}\lrcorner\eta\,.
\end{gather} 
Supersymmetric M2-branes sit at points $p$ where the Hamiltonian function is extremized $\dd h_{M}(p) = 0$, or equivalently 
\begin{gather}\label{eq:zeta prop to Reeb}
	\zeta_{M}\lrcorner\dd{\eta}\big|_p=0\,,\quad\Longrightarrow\quad \zeta_M\big|_p \propto \zeta_R\,.
\end{gather}

It should be noted that the Hamiltonian function may have multiple critical points, each of which can support a supersymmetric M2-brane wrapping the chosen $\UU(1)_{M}$. In the saddle point expansion for the M2-brane partition function we must sum up the contribution of all these critical points 
\be
Z_\text{M2}\approx \sum_{\dd h_{M}(p) = 0} \e^{-S_\text{cl}(p)}Z_\text{1-loop}(p)\,.
\ee 
Although our computations and results are in principle applicable to all saddles, we focus on the leading saddle for which the classical action is smallest. Other saddles will be exponentially suppressed when compared to this leading saddle.

The classical metric on the M2-brane can now be expressed as follows 
\begin{gather}\label{M2 brane metric}
	\dd{s}^{2}_\text{M2}=\frac{R^{2}}{4}\dd s_{\text{AdS}_2}^2+R^{2}c^2\dd\psi^{2}\,,
\end{gather}
where $\psi$ parametrizes the M-theory circle which when pulled back to the M2-brane worldvolume has radius $Rc$ where $c=h_{M}(p)$ is the on-shell value of the Hamiltonian function at its critical point. Note that we have calibrated the coordinate $\psi$ to always be $2\pi$-periodic while still reproducing the correct length of the M-theory circle as predicted by supersymmetry. The latter is given in terms of the contact one-form $\eta$ of the underlying Sasaki-Einstein manifold we study as follows
\begin{gather}
	R\int_{S^{1}_{M}}\eta=2\pi R c\,.
\end{gather}
The regularized on-shell action of the M2-brane is now easily computed \cite{Farquet:2013cwa}
\be
S_\text{cl}(p) = \f{2\pi R^{3}}{4 (2\pi\ell_p)^3} (-2\pi) ( 2\pi c) = -2c \mu\,,\qquad \mu =\f{L^3}{\ell_p^3} = \f{R^3}{8\ell_p^3}\,, 
\ee 
where we used that the regularized area of AdS$_2$ is $-2\pi$. This result shows that the classical action of the M2-brane, and therefore the exponential scaling of the dual Wilson loop vev, is controlled only by the Hamiltonian function $h_M$. The largest critical point of $h_M$ gives rise to the dominant saddle. This was successfully compared to the matrix model in \cite{Farquet:2013cwa} and we will come back to this comparison after we have studied the one-loop fluctuations of the M2-brane around its classical saddle.

\subsection{The quadratic action}
Next, let us turn to the main study of this paper which is the one-loop M2-brane partition function around the classical configuration reviewed above. To this end we consider the quadratic fluctuations of the M2-brane worldvolume fields around the classical configuration. These consist of eight real (physical) scalar fields, and eight fermions. The quadratic action is obtained by splitting up the eleven-dimensional tangent bundle into the worldvolume tangent bundle and its normal bundle. We use indices $a=1,2,\dots,8$ to denote the normal bundle directions. The Green-Schwarz action of the eleven-dimensional M2-brane in \cite{Bergshoeff:1987cm} is then expanded to quadratic order in the fields 
\begin{gather}
	S_{\text{M2}}=S_{\text{M2}}^{\text{cl}}+S_{\text{M2}}^{(2)}+\dots
\end{gather}
This expansion can be performed for general backgrounds as explained in \cite{Astesiano:2024sgi} and references therein, and here we simply quote the final result 
\begin{equation}\label{quad action M2}
	\begin{split}
		S_{\text{\tiny{M2}}}^{(2)}&=\frac{1}{(2\pi)^{2}\ell_{p}^{3}}\int\text{vol}_{3}\big( \mathcal{L}_{\text{bos}} + \mathcal{L}_{\text{ferm}} \big)\,,\\
		\mathcal{L}_{\text{bos}}&=-\frac{1}{2}(\tensor{\hat{D}}{^{a}_{b}}\zeta^{b})^{2}+\frac{1}{2}(\tensor{\hat{R}}{^{i}_{aib}}+\tensor{\hat{K}}{_{a}^{ij}}\tensor{\hat{K}}{_{bij}}-\frac{i}{3!}\epsilon^{ijk}\nabla_{a}G_{bijk}+A_{Gica}\tensor{A}{_G^{ic}_b})\zeta^{a}\zeta^{b}\,,\\
		\mathcal{L}_{\rm{ferm}}&=2i\bar{\theta}\Big(\Gamma^i\partial_i+\frac{1}{4}\Gamma^{i}\tensor{\hat\Omega}{_{i}^{AB}}\Gamma_{AB}+\frac{1}{8}\slashed{G}-\f18 \Gamma^i \slashed{G}\Gamma_i\Big)\theta\,,
	\end{split}
\end{equation}
with 
\begin{gather}\label{quad action M2 p2}
	\tensor{\hat{D}}{^{a}_{b}}\zeta^{b}=\nabla \zeta^{a}+\tensor{A}{^a_b}\zeta^{b}\,,\quad  \tensor{A}{_i^{ab}}=\tensor{\hat\Omega}{_{i}^{ab}}+\tensor{A}{_{Gi}^{ab}}\,,\quad \tensor{A}{_{Gi}^{ab}}=-\frac{i}{4}\epsilon_{ijk}G^{abjk}\,.
\end{gather}
Here the dynamical fields are the eight real scalars $\zeta^a$ and the eleven-dimensional (i.e. 32 component) spinor $\theta$. The latter is subject to the $\kappa$-symmetry gauge choice $i\Gamma_{(3)}\theta = -\theta$ where $\Gamma_{i}$ are the eleven-dimensional gamma matrices pulled-back to the M2-brane worldvolume, $\Gamma_{(3)}=\f1{3!}\epsilon^{ijk}\Gamma_{ijk}$, and $\epsilon^{ijk}$ is the three-dimensional Levi-Civita symbol. The gauge fixing condition reduces the number of independent components of the spinor from 32 to 16. In practice we must split the eleven-dimensional spinor into eight three-dimensional spinors, each of which has two components. However, this is best done case-by-case. The remaining objects appearing in \eqref{quad action M2} and \eqref{quad action M2 p2} are geometric quantities that are inherited from the background geometry at the classical M2-brane position. In particular, $\hat K_{aij}$ is the extrinsic curvature, $\tensor{\hat R}{^{i}_{ajb}}$ is the eleven-dimensional Riemann tensor, and $\tensor{\hat\Omega}{_{i}^{AB}}$ is the eleven-dimensional spin connection. When eleven-dimensional objects appear with tangent-bundle or normal-bundle indices, they have been contracted with the appropriate matrices to pull them back to the relevant bundles. 

Once the quadratic action has been determined, the one-loop partition function is computed in the usual way
\begin{gather}\label{formalpathintegral}
	Z_{1-\rm{loop}} \equiv \e^{-\Gamma_{\rm{M2}}}=\int\qty[D\zeta D\theta D\bar\theta] \e^{-S_{\rm{M2}}^{(2)}}\,.
\end{gather} 
Since, by definition, the action is quadratic in the fields, the above path integral reduces to the determinant of the relevant quadratic operators.

Using the classical M2-brane configuration discussed in the last subsection, we now apply \eqref{quad action M2} to obtain a quadratic action for the fluctuating worldvolume fields for a general AdS$_4\times \text{SE}_7$ background. It turns out that the result is remarkably simple and the three-dimensional masses of the fluctuations are given by a universal expression. The main reason for this fact is that even though a priori the M-theory circle is chosen as one of the isometry directions within the KE$_6$ space, on-shell the tangent vector associated with the M-theory circle is proportional to the Reeb vector. This means that the eight normal directions away from the worldvolume consist of two directions in AdS$_4$ that are transverse to the AdS$_2$ part of the worldvolume and six directions along the KE$_6$ space. This allows us to evaluate the terms associated to the three-dimensional mass in \eqref{quad action M2} universally without first choosing the SE$_7$ manifold we would like to consider. In particular it is easy to verify that the extrinsic curvature vanishes and the Riemann tensor factorizes into components on AdS$_{4}$ and SE$_{7}$. The components of the four-dimensional Riemann tensor are easily worked out starting from an explicit metric on AdS$_4$. However, a priori the curvature tensor on SE$_{7}$ is far more involved and is case-specific. Fortunately, the only components we need are $\tensor{R}{^{0}_{a0b}}$, where we momentarily denote the Reeb vector direction with the index 0 and $a,b$ with directions along KE$_6$. This component is worked out in Appendix \ref{app:SE} and is equal to $\delta_{ab}$. 

Using similar techniques we can work out the required components of the four-form flux $G_{4}$, which we recall only has legs along the AdS$_{4}$ part of the geometry. The only term remaining to be analysed is the eleven-dimensional spin connection  $\tensor{\hat{\Omega}}{_{i}^{ab}}$, whose components along AdS$_{4}$ vanish on-shell, while it has non-trivial components on SE$_{7}$, when evaluated on-shell. In other words, the only part of the pull-backed spin connection that contributes is the $\tensor{\hat{\Omega}}{_{\psi}^{ab}}$ matrix, where we fixed the worldvolume index  to be along the M-theory circle direction and which has its $a,b$ legs along the KE$_{6}$ directions. The resulting matrix is an antisymmetric $6\times6$ matrix, which can be parametrized in terms of its eigenvalues $\pm i q_{l}$, with $l=1,2,3$. %As can be inferred 
As suggested from our notation and as we will discuss in more detail below, these eigenvalues appear as gauge charges for the three-dimensional fluctuations along KE$_{6}$. In order to be consistent with the $\kappa$-symmetry gauge choice introduced above, their sum satisfies 
\begin{gather}\label{sum of q}
	\sum_{l=1}^{3}q_{l}=-1\,.
\end{gather}

Let us integrate by parts such that the quadratic action can be expressed as
\begin{gather}\label{string quad action}
	S^{(2)}_{\text{M2}}= \frac{1}{(2\pi)^{2}\ell_{p}^{3}}\int\text{vol}_{3}(\zeta^{a}\hat{\cal K}_{ab}\zeta^{b}+\bar\theta^{a}\hat{\cal D}_{ab}\theta^{b})\,.
\end{gather}
The kinetic operators $\hat{\cal K}_{ab}$ and $\hat{\cal D}_{ab}$ for the worldvolume scalars and fermions respectively, are diagonal in the $ab$-indices and each entry is given by massive Klein-Gordon and Dirac operators on AdS$_2 \times S^1_{M}$
\begin{equation}	
	\hat{\cal K}_{(q,\hat{M})} = -\hat D_{(q)}^2+  \hat{M}^2\,,\qquad
	\hat{\cal D}_{(q,\hat M)} = i \hat{\slashed D}_{(q)} +\hat{M}\,.
\end{equation}
Note that the covariant derivative on the worldvolume is supplemented by a fixed gauge potential $\hat{A}$ along $S^1_{M}$ under which various fields are charged. We therefore write $D_{(q)}=\hat{\nabla} - i q\hat{A}$ where $\hat{A}=c\,\dd\psi$. Thus, in order to fully specify the spectrum we only have to list the charges $q$ and masses $\hat M$ for the fields. These are summarized in Table \ref{tab: 3d spectrum} from which we can read off that all fermions are massless and all bosons have mass equal to $\hat{M}^2R^2=-1$. We also note that each positively charged field is accompanied by a negatively charged field, both of which have charges that are of the same magnitude $\abs{q}$. This explains the even multiplicity of the bosonic and fermionic fields.

\begin{table}[h!]
	\begin{center}
		{\renewcommand{\arraystretch}{1.3}
			\begin{tabular}{@{\extracolsep{10 pt}}llcc}
				\toprule
				Field & d.o.f. & $ |q| $         & $\hat{M}^2R^2$  \\\midrule
				Scalars & $2$ & $3$ &  $ -1$ \\
				&$2$ & $\abs{q_{1}}$  & $-1$ \\
				&$2$ & $\abs{q_{2}}$  & $-1$ \\
				&$2$ & $\abs{q_{3}}$  & $-1$ \\\midrule
				Fermions & 	$2$ & $2$ &  $0$  \\
				& 	$2$ & $\abs{1-q_{1}}$ &  $0$  \\
				& 	$2$ & $\abs{1-q_{2}}$ &  $0$  \\
				& 	$2$ & $\abs{1-q_{3}}$ &  $0$  \\
				\bottomrule
		\end{tabular}}
		\caption{Spectrum of bosonic and fermionic fluctuations.}
		\label{tab: 3d spectrum}
	\end{center}
\end{table}

Until now, we have written everything in the language of the three-dimensional field theory living on the M2-brane. Our next task is to compute the determinants of the kinetic operators which can in turn be assembled into the one-loop M2-brane partition function. To this end we follow the approach of \cite{Giombi:2023vzu}, where the three-dimensional operators were reduced to a tower of two-dimensional ones in order to utilize the well-known Heat kernel method on AdS$_2$. 
We start by separating the derivatives in the AdS$_2$ direction from the direction of the M-theory circle, that is the $\psi$-direction and write the three-dimensional differential operators as
\begin{equation}\label{3d quad action}
	-\f{R^2}{4}\hat D_{(q)}^2 = -\nabla^2 -\f14\Big(\frac{\partial_\psi}{c}-iq\Big)^2\,,\qquad 
	i\f{R}{2}\hat{\slashed D}_{(q)} = i\slashed\nabla +\f{i}{2}\Big(\frac{\partial_\psi}{c}-iq\Big)\sigma_3\,,
\end{equation}
where we used that the three-dimensional gamma matrices are given by the standard $2\times2$ Pauli matrices. We also multiplied the operators by a relevant length scale such that the two-dimensional Laplace and Dirac operators which appear on the right hand side, without hats, denote the operators on the unit-radius AdS$_2$.
Next, we expand the three-dimensional fields into Fourier modes along $\psi$ resulting in a tower of fields with kinetic operators 
\begin{equation}\label{reductionto2D}
	\begin{split}
		\f{R^2}{4}{\cal K}_{(q,n)} &= -\nabla^2-\f14+m_{(q,n)}^2\,,\\
		\f{R}{2}{\cal D}_{(q,n)} &= i\slashed\nabla  - m_{(q,n)}\sigma_3\,,\\
		m_{(q,n)}&=\f12\Big(\frac{n}{c}-q\Big)\,.\\
	\end{split}
\end{equation}
Here we have filled in the masses that appear in Table \ref{tab: 3d spectrum} for the fields, but we have kept the charges arbitrary. We have defined the short-hand expression $m_{(q,n)}$, which depends on the charge $q$ and the level $n$ and appears in both the bosonic and fermionic operators. This quantity can be viewed as the mass of the fermions, but for the scalars we have extracted an explicit factor of $-1/4$ and hence it cannot be regarded as the total scalar mass.  Now, using these results we can construct the effective action of the quantum M2-brane. This is given by summing over all modes of the quantum effective action of the two-dimensional operators of each field, which yields

\begin{equation}\label{eff action M2}
	\begin{split}
		\Gamma_\text{M2} &= \sum_{n=-\infty}^\infty \sum_q \bigg(\Gamma^{(q,n)}_{\text{bos}}-\Gamma^{(q,n)}_{\text{ferm}}\bigg)\,,\\
		\Gamma^{(q,n)}_{\text{bos}}&= \f12\log\det\bigg(\f{R^2}{4}{\cal K}_{(q,n)}\bigg)\,,\\
	\Gamma^{(q,n)}_{\text{ferm}}&= \f12\log\det\bigg(\f{R}{2}{\cal D}_{(q,n)}\bigg)\,,
	\end{split}
\end{equation}
where we have explicitly performed the path integral in \eqref{formalpathintegral} resulting in the operator determinants. Notice that we included the length scales in the determinants. At each level, these control the logarithmic divergences, but upon summing over $n$ the logarithmic divergences cancel \cite{Giombi:2023vzu}.

It is worth noting that the level $n=0$ modes are the string modes. That is, the modes we would encounter if we consider the quadratic fluctuations of the fundamental string in AdS$_{4}$ around its classical configuration. In this case, the ten-dimensional background in which the string propagates is the type IIA supergravity solution AdS$_{4}\times\rm{M}_{6}$, which we obtain by reducing the  AdS$_{4}\times$SE$_{7}$ background over the M-theory circle, see Figure \ref{fig:M-theory circle} and Section \ref{sec:MtoIIA} for more details.

We may now proceed to the evaluation of the scalar and fermion effective actions of \eqref{eff action M2}. To achieve this, we resort to the Heat kernel method and we refer the reader to Appendix \ref{app:HK} for a review of the derivation. The result for the effective actions can be summarized as follows
\begin{equation}\label{QEAforbosferm}
\begin{split}
	\Gamma^{(q,n)}_{\text{bos}} &= \f{1}{24}(1+\log 2) -\f12 \log A +\f12 \int_0^{m_{(q,n)}^2}\psi(\sqrt{x}+1/2)\dd x\,,\\
	\Gamma^{(q,n)}_{\text{ferm}} &= -\f{1}{12}+ \log A +\f12 |m_{(q,n)}| + \f12 \int_0^{m_{(q,n)}^2}\psi(\sqrt{x})\dd x\,,\\
\end{split}
\end{equation}
where $A$ is the Glaisher constant and $\psi$ is the digamma function. 

As implied by the spectrum in Table \ref{tab: 3d spectrum}, it is convenient to group the two scalar fluctuations with charge $\pm q_l$ together with the fermion fluctuation with charges $\pm \abs{1-q_l}$. To this end let us define $q_0 = 3$ to include the first (and the fifth) row  in Table \ref{tab: 3d spectrum} in our discussion. The combined quantum effective action for each $q_l$ is
\be\label{eq:gamma with qs}
\boldsymbol{\Gamma}^{(q_l,n)}=\Gamma^{(q_l,n)}_{\text{bos}}+\Gamma^{(-q_l,n)}_{\text{bos}}+\Gamma^{(|1-q_l|,n)}_{\text{ferm}}+\Gamma^{(-|1-q_l|,n)}_{\text{ferm}}\,.
\ee
Using the explicit expressions in \eqref{QEAforbosferm}, we arrive at
\be
\begin{split}
\boldsymbol{\Gamma}^{(q, n)} = &|1-q+ c^{-1}n |\bigg[\frac{1}{2}+\log\Gamma\Big(\frac{\abs{1-q+ c^{-1}n}}{2}\Big)\bigg] -| q + c^{-1}n|\log\Gamma\Big(\f{1+\abs{q+ c^{-1}n}}{2}\Big)\\
&+2\psi^{(-2)}\Big(\f{1+\abs{q+ c^{-1}n}}{2}\Big)-2\psi^{(-2)}\Big(\f{\abs{1-q + c^{-1}n }}{2}\Big) -\frac{1}{4}(1+2\log 2\pi)\,.
\end{split}
\ee
Now, 
combining the contribution for all four charges, we can rewrite \eqref{eff action M2} as
\begin{gather}\label{eq: M2 eff action sum}
\begin{aligned}
		\Gamma_\text{M2}&= -\sum_{n=-\infty}^\infty \sum_{l=0}^{3}\boldsymbol{\Gamma}^{(q_l,n)}\,.
\end{aligned}
\end{gather}

At this stage the general analysis has run its course as we need the explicit values for $q_l$ in order to continue. As we have discussed, these are sensitive to the SE$_7$ manifold in question. In the following section we study a few specific examples of SE$_{7}$ manifolds, compute the associated charges and work out the sum over $n$ resulting in the full answer for the quantum effective action. 

\subsection{Comparison with field theory}\label{sec:WL prop}
Combining the classical action for the M2-brane with its one-loop partition function, the M2-brane partition function to this order reads
\begin{gather}\label{M2 part funct}
	Z_{\rm{M2}}\approx \e^{-\Gamma_{\text{M2}}}\e^{2c\mu}\,,
\end{gather}
where $\mu = R^3/8\ell_p^3$. In order to interpret this as the Wilson loop expectation value in the dual field theory we identify the supergravity parameters $c$ and $\mu$ with suitable field theory parameters. Usually, we relate $\mu$ to the rank $N$ through flux quantization
\begin{gather}\label{flux quant}
N=-\frac{1}{(2\pi \ell_{p})^{6}}\int_{\text{SE}_{7}}\star_{11}G_{4}\,,
\end{gather}
resulting in
\begin{equation}\label{M and FT params}
\mu^2=\frac{\pi^{6}N}{6\,\text{vol(SE$_{7}$)}}\,.
\end{equation}
However recently \cite{Gautason:2025plx}, it has been argued that the M2-brane partition function does not compute the Wilson loop vacuum expectation value in the canonical ensemble in the dual field theory.\footnote{See also \cite{Cassia:2025aus} for a similar proposal.} This is because the microscopic definition of M2-branes only knows about the three-form potential $A_3$ and not the dual potential $A_6$. To leading order in the supergravity descriptions, these two quantities are related by a Hodge duality, but this is a leading order (classical) result. Fundamentally, observables in M-theory that are computed by quantizing M2-branes are a function of $A_3$, which in the current context means that they are functions of $\mu$ rather than $N$. The implication of this is that when we compare to field theory, the M2-brane partition function computes observables in the grand canonical ensemble. In order to compare to the Wilson loop vev in the canonical ensemble we must perform a Laplace transform
\be\label{laplatrafo}
\langle W(N) \rangle = \f{1}{Z(N)}\f{1}{2\pi i}\int_{C} \dd \mu \, \e^{\mathcal{Z}_{\rm{M2}}(\mu) - \mu N}\langle W(\mu) \rangle\,,
\ee
where $Z(N)$ is the partition function obtained as the Laplace transform of the exponentiated grand canonical potential $\e^{\mathcal{Z}_{\rm{M2}}(\mu)}$. Here $\mathcal{Z}_{\rm{M2}}(\mu)$ denotes the partition function of an M2-brane with compact topology and to leading order is given by the eleven-dimensional on-shell supergravity action $\mathcal{Z}_{\rm{M2}}(\mu) = - S_\text{sugra} +\cdots$ as a function of $\mu$. 

In \cite{Gautason:2025plx} it was noticed that for many examples, the M2-brane partition function is one-loop exact\footnote{This holds for non-degenerate M2-branes, it is unclear if the cubic polynomial in \eqref{grandcanonicalpot} can be explained by the fact that the corresponding M2-brane is in some sense one-loop exact.} implying that observables in the grand canonical ensemble are particularly simple. Indeed, when a Fermi gas formalism is available in the dual field theory, the $S^3$ partition function is given by an Airy function (see \eqref{airy conj}). This is related to the fact that the grand canonical potential $\mathcal{Z}_{\rm{M2}}(\mu)$ (usually denoted by $J(\mu)$ in the three-dimensional QFT literature) is a cubic polynomial plus terms exponentially suppressed in $\mu$
\be\label{grandcanonicalpot}\begin{split}
\mathcal{Z}_{\rm{M2}}(\mu)&=J(\mu)=J^{\rm{p}}(\mu)+J^{\rm{np}}(\mu)\,,	\\
J^{\text{p}}(\mu)&=\frac{\mathcal{C}}{3}\mu^{3}+\mathcal{B}\mu+\mathcal{A}\,.
\end{split}\ee
The non-perturbative terms $J^{\rm{np}}(\mu)$ descend from worldsheet instantons $\mathcal{O}(\e^{-\# \mu/k})$ and membrane instantons $\mathcal{O}(\e^{-\# \mu})$ \cite{Drukker:2011zy}, as well as from combinations of the two exponential behaviours that come from bound states \cite{Hatsuda:2013gj}.

It is important to note that in the large $N$ limit the Laplace transform in \eqref{laplatrafo} reduces to a Legendre transform that for the Wilson loop, simply implements the flux quantization condition and relates $\mu$ to $N$ as in \eqref{M and FT params}. As such, the distinction between the two ensembles does not seem particularly important in the large $N$ limit. We will however speculate that the M2-brane partition function discussed above is one-loop exact and thus we may use the Laplace transform above to give a formula that is perturbatively exact in $N$ for the Wilson loop in the canonical ensemble. To achieve this we must give the cubic grand canonical partition function as an input, and in the examples discussed below we will borrow results from the literature.

To this end, a crucial point is that the quantum effective action which appears in \eqref{M2 part funct} does not depend on $\mu$. Indeed, the only dependence on $\mu$ appears in the classical action. Therefore, the Laplace transform is particularly easy to perform and yields
\begin{gather}\label{WL prop}
\langle W(N)\rangle^{\rm{p}}=\e^{-\Gamma_{\text{M2}}}\frac{Z^{\rm{p}}_{\rm{SCFT}}(N-2c)}{Z^{\rm{p}}_{\rm{SCFT}}(N)}=\e^{-\Gamma_{\text{M2}}}\frac{\text{Ai}(\mathcal{C}^{-1/3}(N-2c-\mathcal{B}))}{\text{Ai}(\mathcal{C}^{-1/3}(N-\mathcal{B}))}\,.
\end{gather} 
Here the parameters $\mathcal{B}$ and $\mathcal{C}$ are the ones that appear in the $S^3$ partition function \eqref{airy conj} and of course also in the grand canonical potential \eqref{grandcanonicalpot}.
It should be noted that in order to obtain the full $1/2-$BPS Wilson loop vacuum expectation value we should add terms of $\mathcal{O}(\e^{-\# \sqrt{N}})$ to this formula. However, these are beyond the scope of this paper. 
Finally, the large $N$ expansion of \eqref{WL prop}, which could have been obtained by the Legendre transform discussed above, reads
\begin{gather}
\langle W\rangle \approx \e^{-\Gamma_{\text{M2}}}\exp\bigg(\frac{2c N^{1/2}}{\mathcal{C}^{1/2}}\bigg)\,.
\end{gather}

\subsection{Reduction to type IIA supergravity}\label{sec:MtoIIA}
We close this section with some remarks on the dimensional reduction to ten-dimensional type IIA supergravity. For clarity, we restrict ourselves to the case where the eleven-dimensional solution is of the form AdS$_{4}\times$SE$_{7}/\mathbf{Z}_{k}$ where the orbifolding acts without fixed points. 

\begin{figure}[h]
	\centering
	\includegraphics[width=13cm]{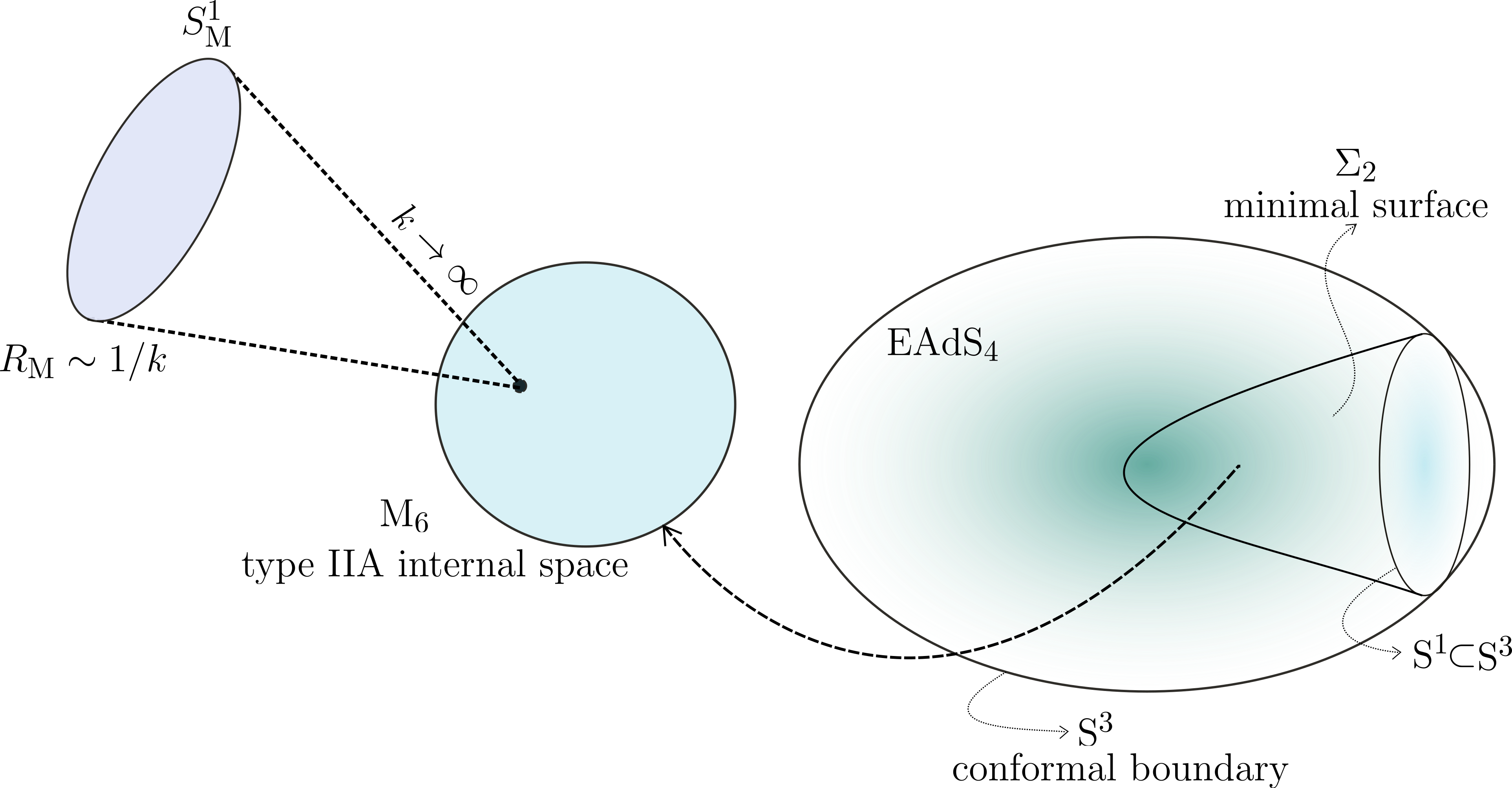}
	\caption{A schematic visualization of the embedding of the M-theory circle $S^{1}_{M}$ in eleven-dimensional supergravity and its dimensional reduction to type IIA supergravity. We observe that in the type IIA limit where $k\rightarrow\infty$, $S^{1}_{M}$ is reduced to a fixed point in the internal manifold M$_{6}$ of the type IIA geometry. On the field theory side the Wilson loop resides on the equator of $S^{3}$. In the dual gravity picture this translates to a holographic Wilson loop that lies on the Hopf circle $S^{1}\subset S^{3}$. This three-sphere is the conformal boundary of Euclidean AdS$_{4}$, inside of which the dual fundamental string occupies the minimal surface $\Sigma_{2}\cong\text{AdS}_{2}\subset$\,AdS$_{4}$, whose conformal boundary is that of $S^{1}$.}
	\label{fig:M-theory circle}
\end{figure} 
The type IIA limit of these backgrounds can be accessed by taking $k,N \rightarrow \infty$ with 't Hooft coupling $\la=N/k=\text{fixed}$ and more generally $k\ll N\ll k^{5}$, while one enters the M-theory regime for $N\gg k^{5}$ with $N\rightarrow\infty$ and $k=\text{fixed}$. Taking the type IIA limit thus reduces the eleven-dimensional AdS$_{4}\times\text{SE}_{7}/\mathbf{Z}_{k}$ background to ten-dimensional AdS$_{4}\times \text{M}_{6}$ by reducing over the M-theory circle. This is visualized in Figure \ref{fig:M-theory circle}. It should be noted that M$_{6}$ should not be confused with the KE$_6$ introduced in \eqref{11d metric} since in general the M-theory circle lies within the latter. We can rewrite the eleven-dimensional metric of \eqref{11d metric} more conveniently as
\begin{gather}\label{eq:11d metric differently}
	\dd{s}^{2}_{11}=R^{2}\qty(\frac{1}{4}\dd{s}_{\text{AdS}_{4}}^{2}+\dd{s}_{\text{M}_{6}}^{2}+\beta^{2}(\dd{\psi}+k\rho)^{2})\,,
\end{gather}
where $\rho$ is a one-form and $\beta$ is a function that in general depends on the coordinates on M$_{6}$.  We  recall that at the classical configuration of the M2-brane, this function satisfies $\beta=h_{M}(p)=c$, where $p\in \text{M}_{6}$ is a critical point of the Hamiltonian function and $c$ is inversely proportional to $k$.

As will be reviewed in Section \ref{sec:famB gravity duals}, in the type IIA limit the Wilson loop operator is dual to a fundamental string, whose minimal surface is $\Sigma_{2}\subset$\,AdS$_{4}$ and has conformal boundary $S^{1}$. From the above, by performing a dimensional reduction of \eqref{eq:11d metric differently} and the worldvolume of the M2-brane along the $\psi$ coordinate one can show that the induced metric on the string worldsheet is given by
\begin{gather}
	\dd{s}_{\text{ws}}^{2}=\frac{cR^{3}}{4\sinh^{2}\s}(\dd{\s}^{2}+\dd{\tau}^{2})\,,
\end{gather}
where the worldsheet coordinates are $(\s,\tau)$ and the on-shell value of the dilaton reads
\begin{gather}
	\e^{\phi_{0}}=g_{s}=(cR)^{3/2}\,.
\end{gather}
Now, as noted in \cite{Giombi:2020mhz} this Wilson loop operator can be written as a genus expansion in the small string coupling $g_{s}$ as follows 
\begin{gather}\label{WL genus expansion}
	\langle W\rangle=\sum_{p=0}^{\infty}a_{p}\qty(\frac{\sqrt{T}}{g_{s}})^{1-2p}\e^{2\pi T}(1+\mathcal{O}(T^{-1}))\,,
\end{gather} 
where the effective tension is $T=\mu c/\pi$.
Following this logic and in similar fashion as was noted for the case of ABJM in \cite{Giombi:2023vzu}, we argue that the M2-brane partition function in \eqref{M2 part funct} should reproduce the resummed genus expansion in the large tension $T$ limit. 
The coefficients $a_{p}$ can be extracted from \eqref{eq: M2 eff action sum} and are in general functions of the charges $q_{l}$. For the case of the 1/2-BPS Wilson loop in ABJM theory, computed originally in \cite{Klemm:2012ii} and discussed in the context of M2-branes in \cite{Giombi:2023vzu}, these coefficients satisfy the following relation,
\begin{gather}\label{a_p coeff}
	a_{p}=\frac{2(4^{p}-2)\zeta(2p)}{(8\pi)^{p}}\,, \qquad p\geq1\,,
\end{gather}
with $a_{0}=0$. In the examples we study in the section below, we will show that this relation also holds for the holographic Wilson loop of $Q^{1,1,1}/\mathbf{Z}_{N_{f}}$, as well as for the ADHM theory. In the latter case however, we find that $a_{0}=1/2$. For other cases, such as $V_{5,2}$, discussed below we find a much more complicated expression than \eqref{a_p coeff}.

\section{Examples}
\label{sec:examples famA}
In this section we consider a number of explicit examples for a known pair of Sasaki-Einstein manifold and its dual ${\cal N}=2$ SCFT.
Specifically, we will study the gravity duals of the ABJM, ADHM, $Q^{1,1,1}$, $Q^{2,2,2}$, $V_{5,2}$ and $M^{3,2}$ theories. We highlight once more that the corresponding holographic dual field theories in general are given by a $\prod_{i=1}^{\mathcal{G}}U(N)_{k_{i}}$ Chern-Simons quiver whose Chern-Simons levels satisfy
\begin{gather}
	\sum_{i=1}^{\mathcal{G}}k_{i}=0\,.
\end{gather} 

For notational convenience in the following examples we will only report the Chern-Simons level vector $\boldsymbol{k}=(k_{1},k_{2},\dots,k_{\mathcal{G}})$ that is associated to the respective field theory dual we consider, together with the corresponding matter content.  Moreover, we note that for each dual geometry we make a specific choice for the M-theory circle, which in the language of the field theory duals means making a specific choice for the holographic Chern-Simons-matter theory. In the examples that follow it is sufficient to report the seven-dimensional SE metric, which allows us to write down the Hamiltonian function $h_{M}$, the radius of the M-theory circle $c$, and the value of the unspecified charges $q_{l}$. From there we can implement our above results to obtain a prediction for the vacuum expectation value of the WL. 
\subsection{ABJM}
\label{sec:Example ABJM}
\begin{figure}[h]
	\centering
	\includegraphics[width=5cm]{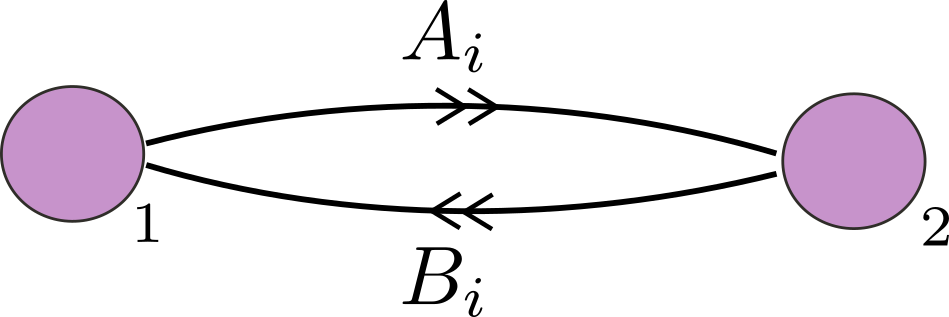}
	\caption{Quiver diagram of ABJM.}
	\label{fig:quiver ABJM}
\end{figure}

The first example we consider is the dual to ABJM theory \cite{Aharony:2008ug}, given by AdS$_{4}\times S^{7}/\mathbf{Z}_{k}$ in eleven-dimensional supergravity. The M2-brane partition function dual to the supersymmetric WL has previously been studied in \cite{Giombi:2023vzu} but for completeness we repeat the analysis here. The ABJM theory is an $\mathcal{N}=6$ Chern-Simons-matter theory with $\boldsymbol{k}=(k,-k)$ and has enhanced $\mathcal{N}=8$ supersymmetry when $k=1,2$. It consists of two pairs of chiral fields: $A_{i}$ and the conjugate $B_{i}$, with $i=1,2$, that transform in the bifundamental representations of the two gauge groups, i.e. $(N,\bar{N})$ and $(\bar{N},N)$ respectively. This gives rise to the quiver shown in Figure \ref{fig:quiver ABJM}. 

Now, the metric of the eleven-dimensional dual geometry is given by \eqref{11d metric} with the $S^{7}/\mathbf{Z}_{k}$ metric given by
\begin{gather}\label{metric S7}
	\dd{s}_{S^{7}}^{2}=\dd{s}_{\mathbf{CP}^{3}}^{2}+\eta^{2}\,,
\end{gather}
where the Kähler-Einstein metric reads
\begin{gather}
	\begin{aligned}
		\dd{s}_{\mathbf{CP}^{3}}^{2}=\sum_{i=1}^{6}(e^{i})^{2}\,,
	\end{aligned}
\end{gather}
with frames $e^{i}$ 
given by 
\begin{gather}
	\begin{aligned}
		&e^{1}=\dd\theta\,,\quad e^{2}=\frac{1}{2}\sin\theta\dd\theta_{1}\,,\quad
		e^{3}=\frac{1}{2}\sin\theta\sin\theta_{1}\dd\phi_{1}\,,\\ &e^{4}=\frac{1}{2}\cos\theta\dd\theta_{2}\,,\quad
		e^{5}=\frac{1}{2}\cos\theta\sin\theta_{2}\dd\phi_{2}\,,\\ &e^{6}=\frac{1}{2}\sin\theta\cos\theta(2\dd{\varphi}+\cos\theta_{1}\dd\phi_{1}-\cos\theta_{2}\dd\phi_{2})\,.
	\end{aligned}
\end{gather}
In these coordinates we have 
\begin{gather}
	\eta=\dd{y}+\frac{1}{2}(\sin^{2}\theta\cos\theta_{1}\dd{\phi_{1}}+\cos^{2}\theta\cos\theta_{2}\dd\phi_{2}-\cos 2\theta \dd\varphi)\,,
\end{gather}
with coordinate ranges $\theta\in[0,\pi/2]$, $\theta_{1,2}\in[0,\pi]$, $\{\phi_{1,2},y\}\in[0,2\pi)$ and $\varphi\in[0,4\pi/k)$.\footnote{It is easy to show that this choice is equivalent to orbifolding $y$ instead, such that $\varphi\in[0,4\pi)$ and $y\in[0,2\pi/k)$.}

From the above we find that the direction of the M-theory circle is identified with $\varphi$ and hence the Hamiltonian function is given by\footnote{Recall that the Hamiltonian function satisfies $\int_{S^{1}_{M}}\eta=2\pi h_{M}({p_{c}})$ and hence is normalized in such a way that the M-theory angle has periodicity $2\pi$. In this example we have $\varphi\sim\varphi+4\pi/k$ and thus the Hamiltonian function is ``rescaled'' by a factor of $2/k$.} 
\begin{gather}
	h_{M}=\frac{1}{k}\cos 2\theta\,,	
\end{gather}
with critical values $\theta=\{0,\pi/2\}$. 
This function is maximized at $\theta=0$, which we identify with the position of the leading M2-brane saddle. Then the charges are $q_{l}=\{3,1,-1,-1\}$, while the radius $c$ and the SE$_{7}$ volume in this example are given by
\begin{gather}\label{vol S7}
	c=\frac{1}{k}\,,\qquad 	\text{vol}(S^{7}/\mathbf{Z}_{k})=\frac{\pi^{4}}{3k}\,.
\end{gather}
Inserting this information into (\ref{eq:gamma with qs} - \ref{eq: M2 eff action sum}) yields the following answer for the effective action of the M2-brane when $k>2$
\begin{gather}\label{eff action ABJM k>2}
	\Gamma_{\text{M}2}=\sum_{n=1}^{\infty}\qty(2\log\frac{kn}{2}+\log\Big(1-\frac{4}{k^{2}n^{2}}\Big))\,,
\end{gather}
which can be easily evaluated using $\zeta$-regularization, that is 
\begin{gather}
	\zeta(0)=\sum_{n=1}^{\infty}1=-\frac{1}{2}\,,\qquad \zeta'(0)=\sum_{n=1}^{\infty}\log n=\frac{1}{2}\log2\pi\,,
\end{gather}
and Euler's identity 
\begin{gather}
	\log\frac{\sin(\pi x)}{\pi x}=\sum_{n=1}^{\infty}\log\Big(1-\frac{x^{2}}{n^{2}}\Big)\,.
\end{gather}
Using the above tools, we can evaluate \eqref{eff action ABJM k>2} yielding the one-loop M2-brane partition function \cite{Giombi:2023vzu}
\begin{gather}
	\e^{-\Gamma_{\text{M2}}}=\frac{1}{2\sin(\frac{2\pi}{k})}\,, \qquad k>2\,.
\end{gather}

For $k=1,2$ the above calculation does not hold and we have to compute the partition function separately. For $kn=1$ we find
\begin{gather}\label{eff action ABJM kn=1}
	\sum_{l=0}^{3}\qty(\boldsymbol{\Gamma}^{(q_{l},1)}+\boldsymbol{\Gamma}^{(q_{l},-1)})\Big\rvert_{k=1}=\log\frac{2}{3}\,,
\end{gather}
while the case of $kn=2$ yields
\begin{gather}\label{eff action ABJM nk=2}
	\sum_{l=0}^{3}\qty(\boldsymbol{\Gamma}^{(q_{l},1)}+\boldsymbol{\Gamma}^{(q_{l},-1)})\Big\rvert_{k=2}=\sum_{l=0}^{3}\qty(\boldsymbol{\Gamma}^{(q_{l},2)}+\boldsymbol{\Gamma}^{(q_{l},-2)})\Big\rvert_{k=1}=\log\pi\,.
\end{gather}
Combining this with the other terms in \eqref{eff action ABJM k>2} we find \cite{Giombi:2023vzu}
\be\label{eq:ZM2 ABJM NF=1,2}
\e^{-\Gamma_{\text{M2}}^{k=1}} = \f14\,,\qquad \e^{-\Gamma_{\text{M2}}^{k=2}} = 1\,.
\ee
As expected and as noted in \cite{Gautason:2025plx}, the prediction for the exact perturbative answer of the vacuum expectation value of the Wilson loop \eqref{WL prop}, with $c=1/k$, \eqref{vol S7} and the parameters of Table \ref{tab:airy param} matches the field theory result of \cite{Klemm:2012ii}\footnote{The answer matches up to a factor of $1/2$, which is due to a different normalization convention employed in \cite{Klemm:2012ii}.}
\begin{gather}\label{ABJM WL pert}
	\langle W\rangle^{\rm{p}}(N,k)=\frac{1}{2\sin(\frac{2\pi}{k})}\frac{\text{Ai}\Big(\tfrac{\pi^{2/3}k^{1/3}}{2^{1/3}}(N-\tfrac{7}{3k}-\tfrac{k}{24})\Big)}{\text{Ai}\Big(\tfrac{\pi^{2/3}k^{1/3}}{2^{1/3}}(N-\tfrac{1}{3k}-\tfrac{k}{24})\Big)}\,,\qquad k>2\,.
\end{gather} 

\subsection{ADHM}
Here we briefly study the vacuum expectation value of the Wilson loop of ADHM theory or sometimes referred to as the $N_{f}$ model. This is an $\mathcal{N}=4$ Chern-Simons-matter theory, which in the language of $\mathcal{N}=2$ supersymmetry contains $N_{f}$ pairs of anti-fundamental and fundamental chiral multiplets $\tilde{A}_{i}$ and $\tilde{B}_{i}$ respectively, with $i=1,\dots,N_{f}$, together with three adjoint chiral multiplets $\Phi_{a}$, with $a=1,2,3$.

\begin{figure}[h]
	\centering
	\includegraphics[width=6cm]{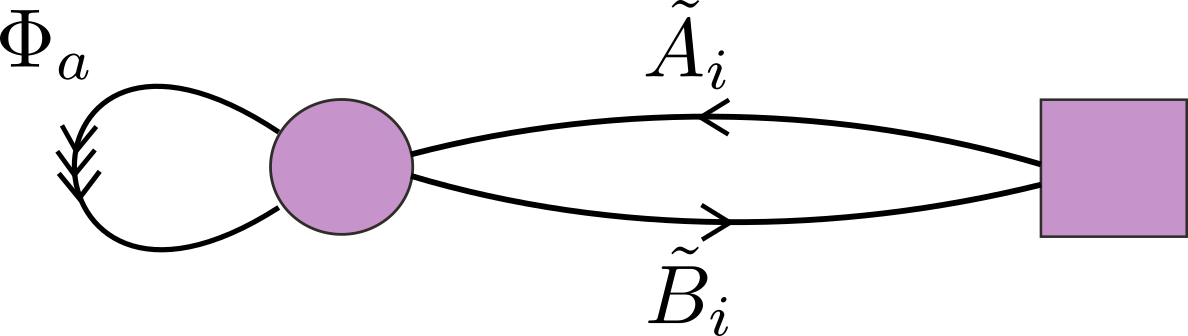}
	\caption{Quiver diagram of ADHM.}
	\label{fig:ADHM quiver}
\end{figure} 

The holographic dual geometry to ADHM is given by AdS$_{4}\times S^{7}/\mathbf{Z}_{N_{f}}$. The SE$_{7}$ metric of the background in this case is given by \eqref{metric S7} where the direction of the M-theory circle is now identified with the angle $\phi_{2}$, which has periodicity
\begin{gather}
	\phi_{2}\sim\phi_{2}+\frac{4\pi}{N_{f}}\,.
\end{gather}
The remaining angles have now the following ranges: $\theta\in[0,\pi/2]$, $\theta_{1,2}\in[0,\pi]$, $\{\varphi,\phi_{1},y\}\in[0,2\pi)$ and the Hamiltonian function in this case reads
\begin{gather}
	h_{M}=\frac{1}{N_{f}}\cos^{2}\theta\cos\theta_{2}\,,
\end{gather}
with three critical values at the points $\theta=\pi/2$ and $\{(\theta,\theta_{2})\,|\, \theta=0\,,\, \theta_{2}=\{0,\pi\}\}$. The Hamiltonian function in this case reaches its maximum value at $\theta=0=\theta_{2}$. Thus, the radius of the M-theory circle and the volume of the SE manifold for this example are given by
\begin{gather}
	c=\frac{1}{N_{f}}\,,\qquad 	\text{vol}(S^{7}/\mathbf{Z}_{N_{f}})=\frac{\pi^{4}}{3N_{f}}\,.
\end{gather}
This specific calibration of the M-theory circle leads to the following result for the charges: $q_{l}=\{3,-1,0,0\}$. 
Evaluating (\ref{eq:gamma with qs} - \ref{eq: M2 eff action sum}) for this data and $N_{f}>2$ results in a similar answer as for ABJM, that is
\begin{gather}\label{eff action ADHM Nf>2}
	\Gamma_{\text{M}2}=\sum_{n=1}^{\infty}\qty(2\log\frac{nN_{f}}{2}+\log\Big(1-\frac{4}{n^{2}N_{f}^{2}}\Big))+\log 2\,.
\end{gather}
The sum can be evaluated exactly as for ABJM, resulting in
\begin{gather}
	\e^{-\Gamma_{\text{M2}}}=\frac{1}{4\sin(\frac{2\pi}{N_{f}})}\,, \qquad N_{f}> 2\,.
\end{gather}

This analysis does not hold for $N_f=1,2$ which have to be treated separately. For $nN_{f}=1$ we find
\begin{gather}\label{eff action ADHM Nfn=1}
	\sum_{l=0}^{3}\qty(\boldsymbol{\Gamma}^{(q_{l},1)}+\boldsymbol{\Gamma}^{(q_{l},-1)})\Big\rvert_{N_{f}=1}=\log\frac{4}{3}\,,
\end{gather}
while the case of $nN_{f}=2$ yields
\begin{gather}\label{eff action ADHM nNf=2}
	\sum_{l=0}^{3}\qty(\boldsymbol{\Gamma}^{(q_{l},1)}+\boldsymbol{\Gamma}^{(q_{l},-1)})\Big\rvert_{N_{f}=2}=	\sum_{l=0}^{3}\qty(\boldsymbol{\Gamma}^{(q_{l},2)}+\boldsymbol{\Gamma}^{(q_{l},-2)})\Big\rvert_{N_{f}=1}=\log\pi\,.
\end{gather}
This can be combined with the other terms in \eqref{eff action ADHM Nf>2}, resulting in
\begin{gather}\label{eq:ZM2 ADHM NF=1,2}
	\e^{-\Gamma_{\text{M2}}^{N_{f}=1}}=\frac{1}{4}\,,\qquad \e^{-\Gamma_{\text{M2}}^{N_{f}=2}}=\frac{1}{2}\,.
\end{gather} 
Notice that for $N_{f}=1$ in \eqref{eq:ZM2 ADHM NF=1,2} we recover exactly the same answer as we found for the holographic Wilson loop of ABJM \eqref{eq:ZM2 ABJM NF=1,2} when the Chern-Simons level is $k=1$, as should be the case since ADHM and ABJM are equivalent when $N_{f}=1$ and $k=1$ respectively. For $N_f>1$ the M2-brane partition function is essentially 1/2 that of the ABJM answer (after mapping $N_f$ to $k$).

By implementing \eqref{WL prop} together with the data from Table \ref{tab:airy param} we find that the perturbatively exact part of the Wilson loop vev of ADHM is
\begin{gather}\label{WL ADHM pert}
	\langle W\rangle^{\rm{p}}(N,N_{f})=\frac{1}{4\sin(\frac{2\pi}{N_{f}})}\frac{\text{Ai}\Big(\tfrac{\pi^{2/3}N_{f}^{1/3}}{2^{1/3}}(N-\tfrac{5}{2N_{f}}+\tfrac{N_{f}}{8})\Big)}{\text{Ai}\Big(\tfrac{\pi^{2/3}N_{f}^{1/3}}{2^{1/3}}(N-\tfrac{1}{2N_{f}}+\tfrac{N_{f}}{8})\Big)}\,, \qquad N_{f}>2\,,
\end{gather} 
which matches the result of \cite{Okuyama:2016pwb} where the matrix model of ADHM was studied.

\subsection{\texorpdfstring{$Q^{1,1,1}$ and $Q^{2,2,2}$ }{Q111 \& Q222}}
In this subsection we will study three different field theories corresponding to two different orbifolds of $Q^{1,1,1}$ and one orbifold of $Q^{2,2,2}$. 
The first example we study is a two-node quiver with $N_f$ fundamental matter, while the second and third examples are given by four-node quivers. 
\subsubsection{$Q^{1,1,1}/\mathbf{Z}_{N_{f}}$}
\begin{figure}[h]
	\centering
	\includegraphics[width=5cm]{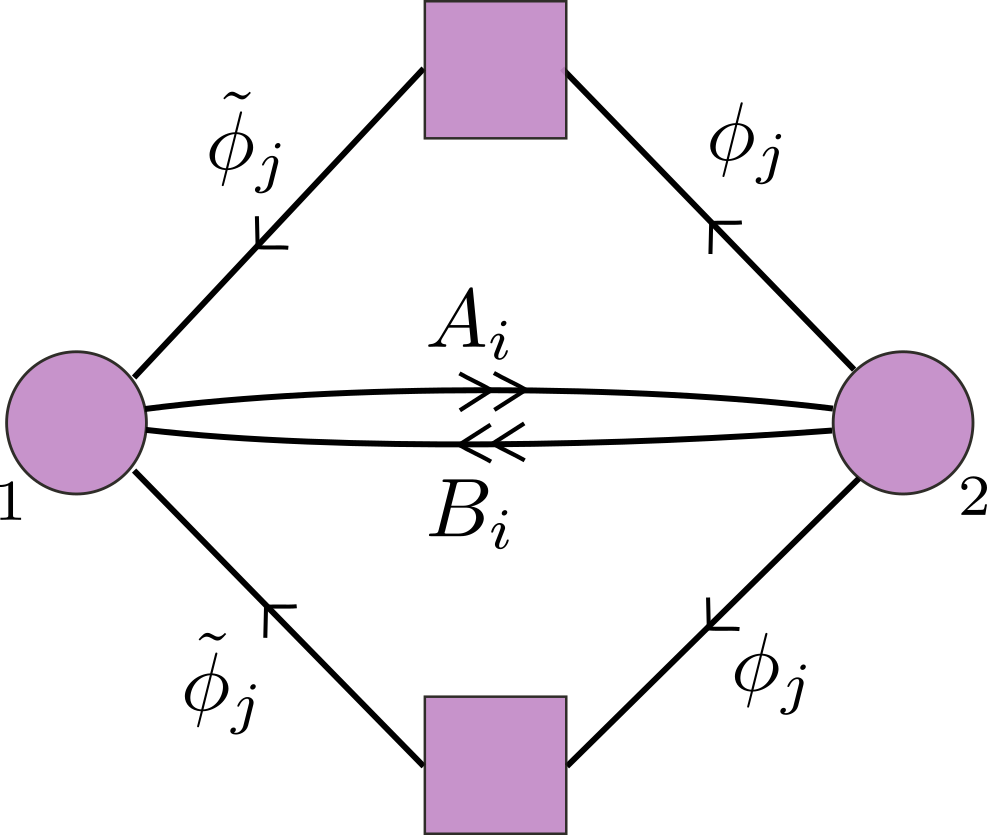}
	\caption{Quiver diagram of $Q^{1,1,1}/\mathbf{Z}_{N_{f}}$.}
	\label{fig:quiver Q111 Nf}
\end{figure}

We start our analysis with the holographic dual to AdS$_{4}\times Q^{1,1,1}/\mathbf{Z}_{N_{f}}$ in eleven-dimensional supergravity. This is a $U(N)\times U(N)$ non-chiral field theory with vanishing CS levels. It consists of two pairs of bifundamental fields $A_{i}$ and the conjugate $B_{i}$ fields with $i=1,2$ that transform in $(N,\bar{N})$ and $(\bar{N},N)$ respectively, as well as $N_{f}$ pairs of fundamental $\phi^{(1)}_{j}, \phi^{(2)}_{j}$ and anti-fundamental $\tilde{\phi}^{(1)}_{j}, \tilde{\phi}^{(2)}_{j}$ chiral multiplets with $j=1,\dots,N_{f}$ \cite{Benini:2009qs}. We summarize this matter content in the quiver diagram of this model in Figure \ref{fig:quiver Q111 Nf}.

The seven-dimensional $Q^{1,1,1}/\mathbf{Z}_{N_{f}}$ metric of interest reads \cite{Nilsson:1984bj}
\begin{gather}\label{eq:metric Q111 Nf}
	\dd{s}^{2}_{7}=\frac{1}{16}\Big(\dd\xi+\sum_{i=1}^{3}\cos\theta_{i}\dd{\phi}_{i}\Big)^{2}+\frac{1}{8}\sum_{i=1}^{3}\Big(\dd{\theta}_{i}^{2}+\sin^{2}\theta_{i}\dd{\phi}_{i}^{2}\Big)\,,	
\end{gather}
where the $S^{2}$ coordinates have ranges $\theta_{i}\in [0,\pi]$, $\phi_{i}\in[0,2\pi)$ and the fibre direction is parametrized by $\xi\in[0,4\pi)$. However, we may define new coordinates $u=(\phi_{1}+\phi_{2})/2$ and  $v=(\phi_{1}-\phi_{2})/2$, each with periodicity $u\sim u+2\pi/N_{f}$ and $v\sim v+2\pi$.\footnote{Equivalently, we could have identified $v\sim v+2\pi/N_{f}$ and $u\sim u+2\pi$, then the M2-brane would be located at $\theta_{1}=0, \theta_{2}=\pi$.} We do this in order to make contact with the known leading order results from \cite{Cheon:2011vi}. The classical solution of the M2-brane was previously found by \cite{Farquet:2013cwa} by identifying the M-theory circle direction with $u$ and the Hamiltonian function is in turn given by
\begin{gather}\label{eq:ham funct a}
	h_{M}=\frac{1}{4N_{f}}(\cos\theta_{1}+\cos\theta_{2})\,,
\end{gather}
whose critical points are located at $\{\theta_{1},\theta_{2}\}=\{0,\pi\}$. The specific choice $\theta_{1}=0=\theta_{2}$ maximizes \eqref{eq:ham funct a} and we identify this with the classical solution of the M2-brane. In this case we find that the geometric quantity $c$ and the volume of the relevant SE manifold respectively read \cite{Farquet:2013cwa}
\begin{gather}\label{Q111c}
	c=\frac{1}{2N_{f}}\,,\qquad 	\text{vol}(Q^{1,1,1}/\mathbf{Z}_{N_{f}})=\frac{\pi^{4}}{8N_{f}}\,.
\end{gather}
Moreover, the charges in this example are exactly the same as the charges of the dual to ABJM, i.e. $q_{l}=\{3,1,-1,-1\}$, hence we can borrow our results from subsection \ref{sec:Example ABJM}. In order to extract the M2-brane partition function for $N_{f}=1$ we should use the $k=2$ results obtained for ABJM because of the explicit factor $1/2$ in \eqref{Q111c}. In this case the effective action of ABJM is trivial, and as such the same holds here for $N_{f}=1$, that is
\begin{gather}\label{eq:Z Nf=1 Q111}
	\e^{-\Gamma_{\text{M2}}^{N_{f}=1}}=1\,.
\end{gather}
By appropriately combining our results, we find that the one-loop partition function for $N_{f}>1$ is given by
\begin{gather}\label{eq:Z Nf Q111}
	\e^{-\Gamma_{\text{M2}}}=\frac{1}{2\sin(\frac{\pi}{N_{f}})}\,,\qquad N_{f}>1\,.
\end{gather} 
It should also be noted that one obtains the same expression for the partition function when taking the orbifold over any appropriate linear combination of $\phi_{i}$ angles. That is, if we define  $u_{1}=(\phi_{1}+\phi_{3})/2$ and  $v_{1}=(\phi_{1}-\phi_{3})/2$ with $u_{1}\sim u_{1}+2\pi/N_{f}$ and $v_{1}\sim v_{1}+2\pi$ or equivalently $u_{2}=(\phi_{2}+\phi_{3})/2$ and  $v_{2}=(\phi_{2}-\phi_{3})/2$ with $u_{2}\sim u_{2}+2\pi/N_{f}$ and $v_{2}\sim v_{2}+2\pi$, we obtain the same answer for the holographic Wilson loop, upon correctly identifying the location of the M2-brane in the transverse space. For completeness we may use the above results together with our conjecture of \eqref{WL prop}, to write down an explicit expression for the Wilson loop. Specifically, we find that for the case of $Q^{1,1,1}/\mathbf{Z}_{N_{f}}$, the perturbative answer of the $1/2$-BPS Wilson loop vacuum expectation value reads
\begin{gather}\label{airy Q111}
	\langle W\rangle^{\rm{p}}(N,N_{f})=\frac{1}{2\sin(\frac{\pi }{N_f})}\frac{\text{Ai}\Big(\left(\tfrac{4\pi^{2}N_{f}}{3}\right)^{\scriptscriptstyle1/3}(N-\tfrac{5}{4N_{f}}+\tfrac{N_{f}}{12})\Big)}{\text{Ai}\Big(\left(\tfrac{4\pi^{2}N_{f}}{3}\right)^{\scriptscriptstyle1/3}(N-\tfrac{1}{4N_{f}}+\tfrac{N_{f}}{12})\Big)}\,,\qquad N_{f}>1\,,
\end{gather}
where we implemented the data of Table \ref{tab:airy param}.

\begin{figure}[h]
	\centering
	\includegraphics[width=5cm]{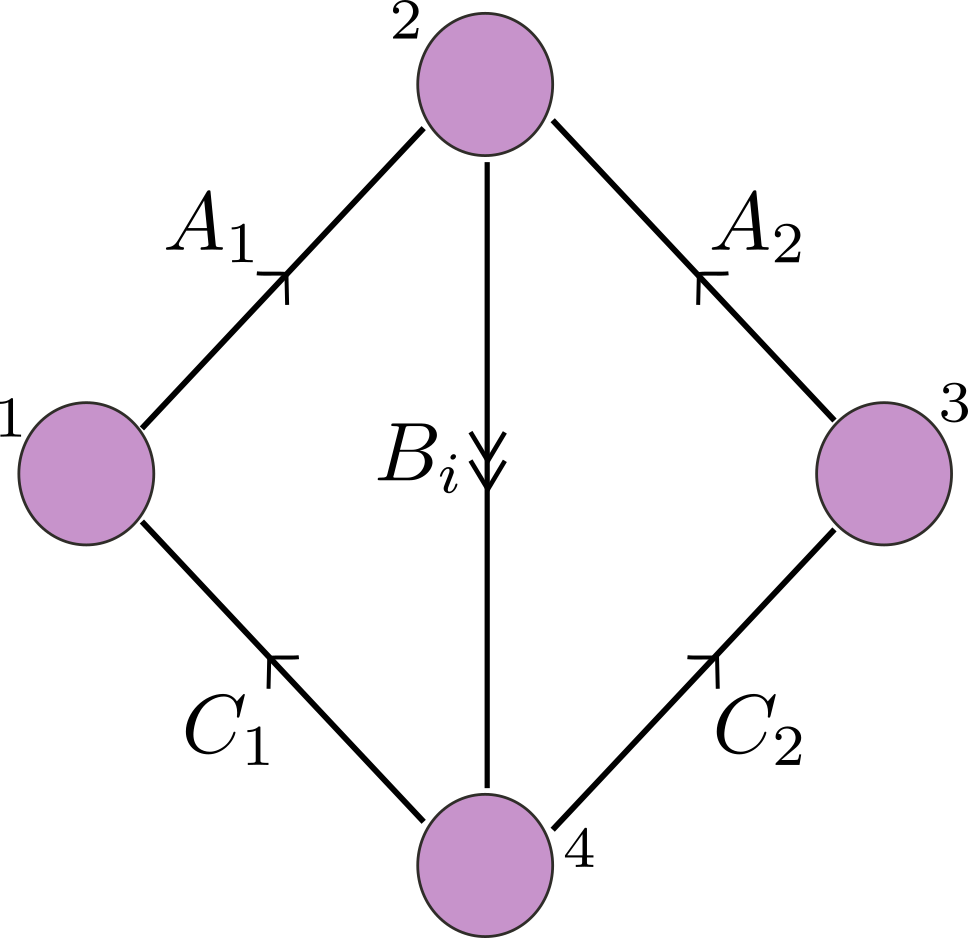}
	\caption{Quiver diagram of $\tilde{Q}^{1,1,1}/\mathbf{Z}_{k}$.}
	\label{fig:quiver Q111}
\end{figure}

Before closing this subsection we note that by replacing $N_{f}\rightarrow k$, this orbifold of the $Q^{1,1,1}/\mathbf{Z}_{N_{f}}$ is also the holographic dual of the four-node quiver gauge theory\footnote{Recently, the matrix model of a similar theory with Chern-Simons levels $\boldsymbol{k}=(k,-k,0,0)$ was studied in \cite{Hosseini:2025jxb}. We thank Seyed Morteza Hosseini for correspondence regarding this model.} with Chern-Simons levels $\boldsymbol{k}=(k,k,-k,-k)$ denoted by $\tilde{Q}^{1,1,1}/\mathbf{Z}_{k}$ in \cite{Franco:2009sp}.\footnote{Please note that we introduced a tilde over the manifold to avoid possible confusions with the $Q^{1,1,1}/\mathbf{Z}_{k}$ manifold we consider in the following subsection, as the two $\mathbf{Z}_{k}$ orbifolds act differently in each case.} Therefore, we can extract the answer for the one-loop M2-brane partition function from (\ref{eq:Z Nf=1 Q111} - \ref{airy Q111}), simply by replacing $N_{f}$ with $k$. 
In this case the non-abelian gauge theory has a $U(1)_{R}\times SU(2)\times U(1)$ symmetry, which matches that of $\tilde{Q}^{1,1,1}/\mathbf{Z}_{k}$.\footnote{As noted in \cite{Franco:2009sp} $Q^{1,1,1}$ has a $U(1)_{R}\times SU(2)^{3}$ symmetry, which is reduced to  $U(1)_{R}\times SU(2)\times U(1)$ due to the orbifold action.} The respective quiver diagram is given in Figure \ref{fig:quiver Q111}. From there one may read off the six chiral fields $A_{1}$, $A_{2}$, $B_{i}$, $C_{1}$, $C_{2}$, with $i=1,2$ that transform in the bifundamental representations of the four $U(N)$ gauge groups, that is $(N,0,\bar{N},0)$, $(0,N,\bar{N},0)$, $(0,0,N,\bar{N})$, $(\bar{N},0,0,N)$, $(0,\bar{N},0,N)$ respectively \cite{Franco:2008um}.

\subsubsection{$Q^{1,1,1}/\mathbf{Z}_{k}$}
We now turn to a different $Q^{1,1,1}/\mathbf{Z}_{k}$ model with Chern-Simons levels $\boldsymbol{k}=(k,0,-k,0)$ and global symmetry $U(1)_{R}\times SU(2)^{2}\times U(1)$ studied in \cite{Aganagic:2009zk,Kim:2012vza}.\footnote{We remark that this theory  may admit a dual description in terms of one of the models that were studied in \cite{Franco:2008um,Franco:2009sp}.} 
In this case, the seven-dimensional $Q^{1,1,1}/\mathbf{Z}_{k}$ metric is given by \eqref{eq:metric Q111 Nf} with $\phi_{1}\in[0,4\pi/k)$ and $\xi \in[0,2\pi)$.\footnote{In what follows, one could replace $\phi_{1}\leftrightarrow \phi_{i}$ and  $\theta_{1}\leftrightarrow \theta_{i}$ with $i=2,3$ respectively, and arrive at the same results.} Similarly as before, the classical solution of the M2-brane is found by identifying the M-theory circle direction with $\phi_{1}$ and the Hamiltonian function for this model is in turn given by 
\begin{gather}\label{eq:ham funct b}
	h_{M}=\frac{1}{2k}\cos\theta_{1}\,,
\end{gather}
whose critical points are located at $\theta_{1}=\{0,\pi\}$ and is extremized when $\theta_{1}=0$. In this case we find that the radius of the M-theory circle $c$ and the volume of the internal manifold take the form
\begin{gather}
	c=\frac{1}{2k}\,,\qquad \text{vol}(Q^{1,1,1}/\mathbf{Z}_{k})=\frac{\pi^{4}}{8k}\,,
\end{gather}
the charges are $q_{l}=\{3,1,1,-3\}$. Plugging this into (\ref{eq:gamma with qs} - \ref{eq: M2 eff action sum}) for  $k>1$ we find that the effective action is given by
\begin{gather}
	\Gamma_{\text{M2}}=\sum_{n=1}^{\infty}2\log(kn) -\log(2\pi)\,.
\end{gather}
Using $\zeta$-regularization we find that the one-loop M2-brane partition function reads
\begin{gather}\label{eq:1-loop other Q111}
	\e^{-\Gamma_{\text{M2}}}=k\,,\qquad k>1\,.
\end{gather} 
In contrast with the analogous results found for the ABJM and ADHM theories, the above expression remains well-behaved for $k=1$.   Therefore, we expect that \eqref{eq:1-loop other Q111} can be analytically continued to this small value of $k$.
Interestingly, this result seems to indicate that the string theory limit  $k\gg 1$ is exact. In other words, only the $p=0$ term in \eqref{WL genus expansion} is non-trivial. It would be interesting to explore what string theory mechanism is responsible for this behaviour. 
\subsubsection{$Q^{2,2,2}/\mathbf{Z}_{k}$}
We can easily generalize the discussion of the previous subsection to the $Q^{2,2,2}/\mathbf{Z}_{k}$ model studied in \cite{Hanany:2008fj,Davey:2009sr,Franco:2009sp,Amariti:2011uw, Kim:2012vza}. This model has Chern-Simons levels  $\boldsymbol{k}=(k,k,-k,-k)$ and the same global symmetry as $Q^{1,1,1}/\mathbf{Z}_{k}$. From the gravity perspective, the difference compared to this latter theory manifests itself in a different periodicity of the angle parametrising the M-theory circle, that is $\phi_{1}\in[0,2\pi/k)$ and of course $\xi \in[0,2\pi)$, as the $Q^{2,2,2}$ manifold results from a $\mathbf{Z}_{2}$ orbifold that acts on $Q^{1,1,1}$ and halves the period of the $\xi$ angle of \eqref{eq:metric Q111 Nf}. Therefore, we can recycle the above results by performing the rescaling $k\rightarrow 2k$, in which case the one-loop M2-brane partition function is given by 
\begin{gather}\label{eq:1-loop Q222}
	\e^{-\Gamma_{\text{M2}}}=2k\,,\qquad k\geq1\,,
\end{gather} 
indicating once again that the string theory limit is exact.
\subsection{\texorpdfstring{$V_{5,2}$}{V52}}\label{sub:v52}
Here we will study both the one-node and two-node quiver theories, which are dual to two different quotients of $V_{5,2}$.

\subsubsection{$V_{5,2}/\mathbf{Z}_{N_{f}}$}
The field theory we will study in this section is the low energy effective field theory on a stack of $N$ M2-branes, whose near-horizon geometry is that of AdS$_{4}\times V_{5,2}/\mathbf{Z}_{N_{f}}$ in weakly curved eleven-dimensional supergravity. This is a $U(N)$ theory with vanishing CS level and three adjoint chiral multiplets $\Phi_{I}$, with $I=1,2,3$ as well as $N_{f}$ pairs of fundamental $\phi_{r}$ and anti-fundamental $\tilde\phi_{r}$ chiral multiplets with $r=1,\dots,N_{f}$ \cite{Jafferis:2009th}. The corresponding quiver is illustrated in Figure \ref{fig:quiver V52 Nf} below.

\begin{figure}[h]
	\centering
	\includegraphics[width=6cm]{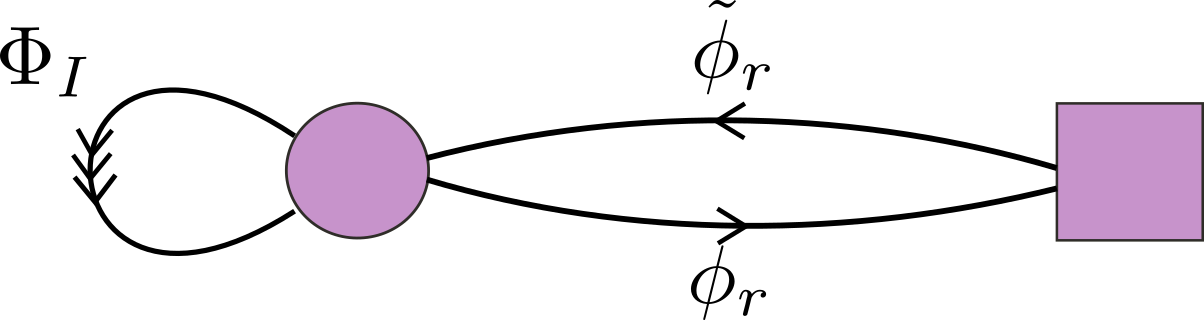}
	\caption{Quiver diagram of $V_{5,2}/\mathbf{Z}_{N_{f}}$.}
	\label{fig:quiver V52 Nf}
\end{figure}

In this case the relevant seven-dimensional SE metric is given by \cite{Bergman:2001qi}
\begin{gather}
	\dd{s}_{7}^{2}=\frac{9}{16}\qty(\dd\xi+\frac{1}{2}\cos\alpha(\dd{\beta}-\cos\theta_{1}\dd\phi_{1}-\cos\theta_{2}\dd\phi_{2}))^{2}+\dd{s}_{\rm{Gr}_{5,2}}^{2}\,,
\end{gather}
where
\begin{gather}
	\begin{aligned}
		\dd{s}_{\rm{Gr}_{5,2}}^{2}=&\frac{3}{32}\Big(4\dd\alpha^{2}+\sin^{2}\alpha(\dd\beta-\cos\theta_{1}\dd\phi_{1}-\cos\theta_{2}\dd\phi_{2})^{2}\\
		&+(1+\cos^{2}\alpha)(\dd\theta_{1}^{2}+\sin^{2}\theta_{1}\dd\phi_{1}^{2}+\dd\theta_{2}^{2}+\sin^{2}\theta_{2}\dd\phi_{2}^{2})\\
		&+2\sin^{2}\alpha\cos\beta\sin\theta_{1}\sin\theta_{2}\dd\phi_{1}\dd\phi_{2}-2\sin^{2}\alpha\cos\beta\dd\theta_{1}\dd\theta_{2}\\
		&+2\sin^{2}\alpha\sin\beta(\sin\theta_{2}\dd\phi_{2}\dd\theta_{1}+\sin\theta_{1}\dd\phi_{1}\dd\theta_{2})\Big)\,,
	\end{aligned}
\end{gather}
with coordinate ranges $\{\phi_{i},\xi\}\in[0,2\pi)$, $\beta\in[0,4\pi)$, $\al\in[0,\pi/2]$, and $\theta_{i}\in[0,\pi)$.
We study the same case that was studied in \cite{Cheon:2011vi} and thus must take the orbifold action along  $u\sim u+2\pi/N_{f}$ where we defined $u=(\phi_{1}+\phi_{2})/2$ and $v=(\phi_{1}-\phi_{2})/2$.\footnote{Again, we note that one could equivalently identify $v\sim v+2\pi/N_{f}$  and $u\sim u+2\pi$, in which case the M2-brane would be situated at $\al=0=\theta_{1}, \theta_{2}=\pi$.}  The corresponding Hamiltonian function is given by
\begin{gather}
	h_{M}=\frac{3}{8k}\cos\al(\cos\theta_{1}+\cos\theta_{2})\,,
\end{gather}
with critical points at $\alpha=\{0,\pi\}$, $\theta_{i}=\{0,\pi\}$ with $i=1,2$. This function takes its largest value when $\al=0=\theta_{i}$. The classical solution of the M2-brane is then given exactly at these points, from which we can extract $c$, which together with the volume of $V_{5,2}/\mathbf{Z}_{N_{f}}$ read 
\begin{gather}\label{c value V52 Nf}
	c=\frac{3}{4N_{f}}\,, \qquad \text{vol}(V_{5,2}/\mathbf{Z}_{N_{f}})=\frac{27\pi^{4}}{128N_{f}}\,.
\end{gather}
Moreover, the charges in this case read $q_{l}=\{3,-1/3,-1/3,-1/3\}$. This yields the following result for the effective action of the M2-brane for $N_{f}>2$
\begin{gather}\label{eff action V52 Nf>2}
	\Gamma_{\text{M}2}=\sum_{n=1}^{\infty}\qty(\log\Big(1-\frac{9}{4n^{2}N_{f}^{2}}\Big)-3\log\qty(\frac{3}{2nN_{f}}\frac{\Gamma(\frac{2+2nN_{f}}{3})}{\Gamma(\frac{1+nN_{f}}{3})}))-3\log\Gamma\Big(\frac{2}{3}\Big)+\log 2\pi\,.
\end{gather}
From \eqref{eff action V52 Nf>2} it is clear that the first term in the summand nicely sums to the logarithm of a sine function as we have seen before. The second term of the summand can be written as a ratio of Barnes double $\Gamma$-functions which are defined as \cite{4531e435-b882-3cb6-9bad-5cab651fc8a8, Bytsko:2006ut}
\begin{gather}
	\log\Gamma_{2}(x | \omega_{1},\omega_{2})=\qty(\pdv{t}\sum_{n_{1},n_{2}=0}^{\infty}(x+n_{1}\omega_{1}+n_{2}\omega_{2})^{-t})\Bigg\lvert_{t=0}\,.
\end{gather}
This results in the following expression for the one-loop M2-brane partition function 
\begin{gather}\label{1-loop Z V52 full}
	\e^{-\Gamma_{\text{M2}}}=\frac{\Gamma(\tfrac{1}{3})^{3}}{4\sin(\tfrac{3\pi}{2N_{f}})}\sqrt{\frac{N_{f}}{3\pi^{3}}}\frac{\Gamma_{2}(\tfrac{2}{3}|\tfrac{2N_{f}}{3},1)^{3}}{\Gamma_{2}(\tfrac{1}{3}|\tfrac{2N_{f}}{3},1)^{3}}\,,\qquad N_{f}>2\,.
\end{gather} 
Once again, we expect that one can analytically continue this result to the case of $N_{f}=1,2$.
Then according to \eqref{WL prop} the full perturbative answer for the WL vev is in this case given by
\begin{gather}\label{WL prop V52/nf}
	\langle W\rangle^{\rm{p}}(N,N_{f})=\frac{\Gamma(\tfrac{1}{3})^{3}}{4\sin(\tfrac{3\pi}{2N_{f}})}\sqrt{\frac{N_{f}}{3\pi^{3}}}\frac{\Gamma_{2}(\tfrac{2}{3}|\tfrac{2N_{f}}{3},1)^{3}}{\Gamma_{2}(\tfrac{1}{3}|\tfrac{2N_{f}}{3},1)^{3}}\frac{\text{Ai}\Big(\tfrac{4\pi^{2/3}N_{f}^{1/3}}{3^{4/3}}(N-\tfrac{29}{16N_{f}}-\tfrac{N_{f}}{48})\Big)}{\text{Ai}\Big(\tfrac{4\pi^{2/3}N_{f}^{1/3}}{3^{4/3}}(N-\tfrac{5}{16N_{f}}-\tfrac{N_{f}}{48})\Big)}\,,
\end{gather}  
for $N_{f}>2$.
On the other hand, by expanding \eqref{eff action V52 Nf>2} around large $N_{f}$, which is the string theory expansion, we obtain\footnote{We are grateful to Evangelos Tsolakidis for assisting us in obtaining this expression.}
\begin{gather}\label{eff action V52 full}
	\begin{aligned}	\Gamma_{\text{M}2}=	&\frac{\sqrt{3}}{2\pi}\sum_{s=1}^{\infty}\frac{(-1)^{s+1}\zeta(2s)}{2^{4s}\pi^{2s}N_{f}^{2s}s}\Big(2\psi^{(2s)}\qty(\tfrac{1}{3})+(2s)!(3^{2s+1}-1)\zeta(2s+1)\Big)\\
	&+\log\Big(\f{4\pi}{\Gamma(\tfrac{2}{3})^3}\sin(\tfrac{3\pi}{2N_{f}})\Big)\,,
	\end{aligned}
\end{gather}
where the first few terms read
\begin{gather}\label{V52 eff action}
\Gamma_{\text{M}2}=\log\Big(\f{6\pi^2}{N_f\Gamma(\tfrac{2}{3})^{3}}\Big)-\frac{7\pi^{2}}{18N_{f}^{2}}-\frac{\pi^{4}}{36N_{f}^{4}}+\mathcal{O}(N_{f}^{-6})\,.
\end{gather}
Using our expansion in \eqref{V52 eff action}, together with the large $N$ limit (keeping $N/N_{f}$ fixed) we can access the string theory limit of our answer
\begin{gather}
	Z_{\text{M2}}=\exp\qty(\frac{4\pi }{3}\sqrt{\frac{N}{N_{f}}}) \qty(\Gamma\Big(\frac{2}{3}\Big)^{3}\frac{N_{f}}{6\pi^{2}}+\mathcal{O}(N_{f}^{-1}))\,, \qquad N_{f}\gg 1\,.
\end{gather}
We note that this result is quite similar to the universal answer for the one-loop string partition function of the gravity duals of family B that we analyse in Section \ref{sec:famB gravity duals}. In fact, the mass spectrum we obtain at the string level (when $n=0$), for this model is the same as the mass spectrum obtained for the universal string operators of Section \ref{sec:famB gravity duals}. For now we close this subsection and provide more comments on this intriguing similarity in the next section.
\subsubsection{ $V_{5,2}/\mathbf{Z}_{k}$}
Now, let us consider the case of  AdS$_{4}\times V_{5,2}/\mathbf{Z}_{k}$, whose field theory dual has been studied in \cite{Martelli:2011qj, Cheon:2011vi}. In this case $V_{5,2}/\mathbf{Z}_{k}$ has isometry group $SU(2)\times U(1)\times U(1)_{R}$, which is identical to the global symmetry group of the dual $\mathcal{N}=2$ Chern-Simons quiver theory with levels $\boldsymbol{k}=(k,-k)$.  The matter content of this theory consists of two chiral fields $A_{i}$, $i=1,2$ transforming in the bifundamental representation, $B_{i}$ transforming in the conjugate bifundamental representation and two massless chiral fields $\Phi_{I}$, $I=1,2$ in the adjoint representations \cite{Martelli:2009ga}. This is depicted in Figure \ref{fig:quiver V52} below. Note that although this field theory is reminiscent of ABJM theory, the chiral fields $\Phi_{I}$ are massless, whereas in the case of ABJM they are massive and can be integrated out at the level of the superpotential in the low energy limit.

\begin{figure}[h]
	\centering
	\includegraphics[width=6.6cm]{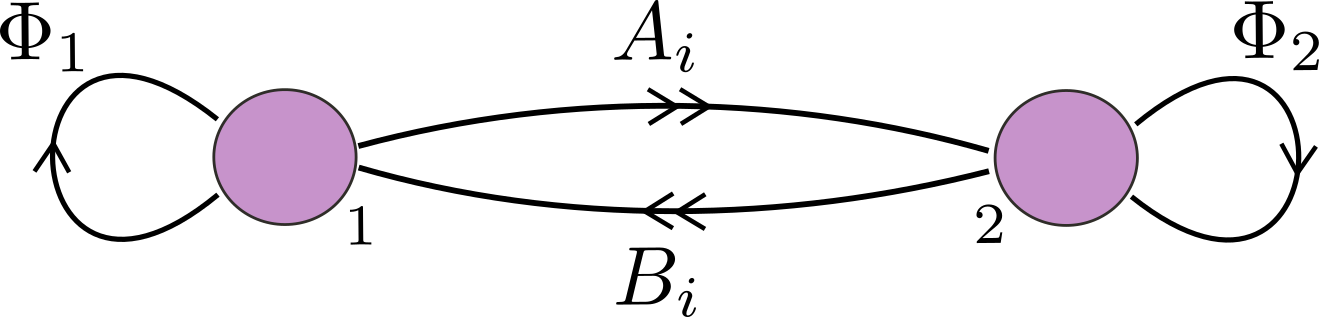}
	\caption{Quiver diagram of $V_{5,2}/\mathbf{Z}_{k}$.}
	\label{fig:quiver V52}
\end{figure}

For this model we study the same case that was studied in \cite{Cheon:2011vi,Martelli:2011qj} and thus take $\beta\sim\beta+4\pi/k$.  The classical solution of the M2-brane is given by identifying the M-theory circle with $\beta$, which yields the following Hamiltonian function 
\begin{gather}
	h_{M}=\frac{3}{4k}\cos\al\,,
\end{gather}
whose critical points are situated at $\alpha=\{0,\pi\}$. It acquires its largest value at $\al=0$, which we identify with the classical solution of the M2-brane. Hence, the radius $c$ and the relevant volume of the internal space are given by 
\begin{gather}\label{c value V52 k}
	c=\frac{3}{4k}\,,\qquad \text{vol}(V_{5,2}/\mathbf{Z}_{k})=\frac{27\pi^{4}}{128k}\,.
\end{gather}
We note that the scalar charges $q_{l}$ are the same as the ones found in the $V_{5,2}/\mathbf{Z}_{N_f}$ model in the previous subsection. As a result, we can recycle all our previous results of the above subsection, by performing the substitution $N_{f}\rightarrow k$. Naturally, the full perturbative part of the vacuum expectation value of the WL will differ compared to \eqref{WL prop V52/nf}, as the $\mathcal{B}$  parameter for this model is not known in the literature. Therefore, the vacuum expectation value of the WL reads
\begin{gather}\label{M2 part funct M32 1}
	\langle W\rangle^{\rm{p}}(N,k)=\frac{\Gamma(\tfrac{1}{3})^{3}}{4\sin(\tfrac{3\pi}{2k})}\sqrt{\frac{k}{3\pi^{3}}}\frac{\Gamma_{2}(\tfrac{2}{3}|\tfrac{2k}{3},1)^{3}}{\Gamma_{2}(\tfrac{1}{3}|\tfrac{2k}{3},1)^{3}}\frac{\text{Ai}\qty(\tfrac{4\pi^{2/3}k^{1/3}}{3^{4/3}}(N-\tfrac{3}{2k}-\mathcal{B}))}{\text{Ai}\qty(\tfrac{4\pi^{2/3}k^{1/3}}{3^{4/3}}(N-\mathcal{B}))}\,,\qquad k>2\,.
\end{gather}

\subsection{\texorpdfstring{$M^{3,2}/\mathbf{Z}_{k}$}{M32/Zk}}
\begin{figure}[h]
	\centering
	\includegraphics[width=5cm]{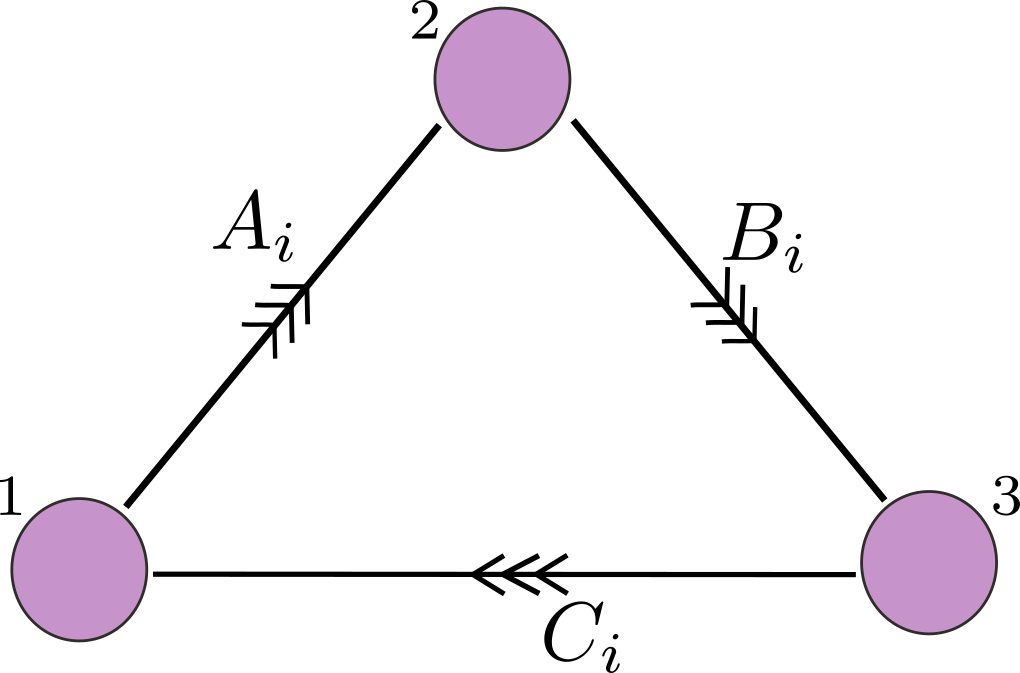}
	\caption{Quiver diagram of $M^{3,2}$.}
	\label{fig:quiver M32}
\end{figure}

Finally, we consider the case of AdS$_{4}\times M^{3,2}/\mathbf{Z}_{k}$, which is dual to the three-dimensional $M^{3,2}/\mathbf{Z}_{k}$ SCFT. This is a three-node quiver gauge theory with CS levels $\boldsymbol{k}=(-2k,k,k)$ \cite{Franco:2009sp}. It has the global symmetry of $U(1)_{R}\times SU(3)\times U(1)$, which matches the isometries of $M^{3,2}/\mathbf{Z}_{k}$. As depicted in Figure \ref{fig:quiver M32} the quiver has three sets of chiral fields $A_{i}$, $B_{i}$ and $C_{i}$ with $i=1,2,3$ that transform in the bifundamental representations of the three gauge groups: $(N,\bar{N},0)$,  $(0,N,\bar{N})$ and  $(\bar{N},0,N)$ respectively \cite{Martelli:2008si}. The matrix model of this Chern-Simons-matter theory was studied in the large $N$ limit by \cite{Amariti:2011uw,Gang:2011jj}, where the free energy on $S^{3}$ was computed. This allowed us to fix the $\mathcal{C}$ parameter for this example in Table \ref{tab:airy param}.

The corresponding gravity dual is equipped with the following seven-dimensional SE metric \cite{Nilsson:1984bj}
\begin{gather}\label{eq:metric M32}
	\begin{aligned}
		\dd{s}_{7}^{2}=&\frac{1}{64}(\dd\xi-3 \sin^{2}\mu \s_{3}+2\cos\theta_{1}\dd\phi_{1})^{2}+\frac{1}{6}(\dd\theta_{1}^{2}+\sin^{2}\theta_{1}\dd\phi_{1}^{2})\\
		&+\frac{3}{4}(\dd\mu^{2}+\frac{1}{4}\sin^{2}\mu(\s_{1}^{2}+\s_{2}^{2}+\cos^{2}\mu\s_{3}^{2}))\,,
	\end{aligned}
\end{gather}
where $\mu\in [0,\pi/2]$, $\theta_{1}$ and $\phi_{1}$ are the polar coordinates on $S^{2}$. The left-invariant $SU(2)$ one-forms read
\begin{gather}
	\begin{aligned}
		\s_{1}=-\sin\psi_{1}\dd\rho_{1}+\cos\psi_{1}\sin\rho_{1}\dd{\varphi}_{1}\,,\quad& \s_{2}=\cos\psi_{1}\dd\rho_{1}+\sin\psi_{1}\sin\rho_{1}\dd\varphi_{1}\,, \\ 
		\s_{3}=\cos\rho_{1}\dd\varphi_{1}&+\dd\psi_{1}\,,
	\end{aligned}
\end{gather}
with $\rho_{1}\in[0,\pi]$, $\varphi_{1}\in[0,4\pi]$, $\psi_{1}\in[0,2\pi)$ and $\phi_{1}\sim \phi_{1}+2\pi/k$. The classical M2-brane solution we consider is given by identifying the coordinate parametrizing the M-theory circle with $\phi_{1}$, and as a result the corresponding Hamiltonian function reads
\begin{gather}\label{eq:ham funct M32}
	h_{M}=\frac{1}{4k}\cos\theta_{1}\,.
\end{gather}
The critical points of this function are given by $\theta=\{0,\pi\}$ and taking $\theta_{1}=0$ yields the maximum value of \eqref{eq:ham funct M32}. With this we find the parameter $c$ and the SE volume
\begin{gather}\label{c value M32}
	c=\frac{1}{4k}\,,\qquad \text{vol}(M^{3,2}/\mathbf{Z}_{k})=\frac{9}{128 k}\pi^{4}\,.
\end{gather}
In this example we find that the scalar charges are $q_{l}=\{3,-3,1,1\}$. Using this together with \eqref{c value M32} yields the following answer for the M2-brane effective action for $k\geq1$ 
\begin{gather}
	\Gamma_{\text{M2}}=\sum_{n=1}^{\infty}2\log(2kn) -\log(2\pi)\,.
\end{gather}
As shown in the previous examples we can easily evaluate these infinite sums using $\zeta$-regularization. 
This
results in the following one-loop M2-brane partition function
\begin{gather}\label{M2 part funct M32}
	\e^{-\Gamma_{\text{M2}}}=2k\,,\qquad k\geq1\,.
\end{gather} 

Using \eqref{M2 part funct M32}, we find that the perturbative answer for the vev of the WL reads
\begin{gather}\label{WL prop M32}
	\langle W\rangle^{\rm{p}}(N,k)=2k\frac{\text{Ai}\qty(\tfrac{4\pi^{2/3}k^{1/3}}{3}(N-\tfrac{3}{4k}-\mathcal{B}))}{\text{Ai}\qty(\tfrac{4\pi^{2/3}k^{1/3}}{3}(N-\mathcal{B}))}\,,\qquad k\geq1\,,
\end{gather}
up to the $\mathcal{B}$ parameter which remains to be determined with a careful analysis of the corresponding matrix model. It should be noted that the leading order exponential behaviour of \eqref{WL prop M32} and thus the leading order answer for the WL in the large $N$ limit, matches the matrix model computation of \cite{Amariti:2011uw,Gang:2011jj} up to a factor of $2/3$, which interestingly is equal with the superconformal $R$-charge $\Delta$.  Perhaps it is not too surprising that this does not match the matrix model analysis to the nose, as this is a chiral theory, which in contrast with non-chiral theories, requires special treatment as highlighted in \cite{Hosseini:2025jxb}. Interestingly, similarly to the $Q^{1,1,1}/\mathbf{Z}_{k}$ example, which was also a chiral theory, this answer is exact in the string theory limit $k\gg 1$.

Before closing this subsection we note that one can study a \textit{different} quiver that the one presented above. This can be achieved by choosing a different M-theory circle. As an example we may consider the case where the coordinate ranges are given by $\phi_{1}\in[0,2\pi)$, $\rho_{1}\in[0,\pi]$, $\varphi_{1}\in[0,4\pi]$ and $\psi_{1}\in[0,2\pi/k)$. Then the SU(3) symmetry is broken and the classical M2-brane solution we consider is given by identifying the direction of the M-theory circle with $\psi_{1}$.  In this case, the corresponding Hamiltonian function reads
\begin{gather}\label{eq:ham funct M32 other}
	h_{M}=\frac{3}{8k}\sin^{2}\mu\,.
\end{gather}
The critical points of this function are given by $\mu=\{0,\pi/2\}$ and taking $\mu=\pi/2$ yields the maximum value of \eqref{eq:ham funct M32 other}. With this we find the parameter $c$ and the SE volume
\begin{gather}\label{c value M32 other}
	c=\frac{3}{8k}\,,\qquad \text{vol}(M^{3,2}/\mathbf{Z}_{k})=\frac{9}{128 k}\pi^{4}\,.
\end{gather}
In this example we find that the scalar charges are $q_{l}=\{3,-5/3,-1/3,1\}$. Using this together with \eqref{c value M32 other} yields the following answer for the M2-brane effective action for $k>1$ 
\begin{gather}
	\Gamma_{\text{M2}}=\sum_{n=1}^{\infty}\qty(2\log\frac{4kn}{3}+\log\qty(1-\frac{9}{16k^{2}n^{2}})-\log\qty(1-\frac{1}{16k^{2}n^{2}})) -\log\frac{2\pi}{3\sqrt{3}}\,,
\end{gather}
which can be evaluated through $\zeta$-regularization. 
This gives the following one-loop M2-brane partition function
\begin{gather}\label{M2 part funct M32 other}
	\e^{-\Gamma_{\text{M2}}}=\frac{4 k\sin(\frac{\pi}{4k})}{3\sqrt{3}\sin(\frac{3\pi}{4k})}\,,
\end{gather} 
which is valid for $k>1$, but as above can possibly be analytically continued to the case of $k=1$.
It is important to note that the dual field theory of this background is not known. Hence, our results provide predictions for future studies.

%----------------------------------------------------------------------------------------
%	Holographic WLs in mIIA
%----------------------------------------------------------------------------------------
\section{Holographic Wilson loop for \famb}
\label{sec:famB gravity duals}
We now turn to the study of theories in family B which are dual to AdS$_{4}$ backgrounds in massive type IIA string theory.
A notable feature of these theories is their direct connection to a class of four-dimensional $\mathcal{N}=1$ quiver gauge theories \cite{Fluder:2015eoa} (see also \cite{Bobev:2021gza}). This connection was inspired by a similar relation that appeared between four-dimensional theories and three-dimensional SCFTs in family A \cite{Martelli:2008si,Aganagic:2009zk}. The four-dimensional theories are referred to as the \textit{parent} theories of the three-dimensional \textit{daughter} field theories and are described by the same quiver diagram and superpotential. For theories in family B, this relation establishes a precise connection between $a$-maximization of the parent theory and $F$-extremization of the daughter theory \cite{Fluder:2015eoa}. Briefly speaking, $a$-maximization involves extremizing the $a$-function with respect to the trial $R$-charges of the matter fields \cite{Intriligator:2003jj}, to obtain an SCFT. Similarly, $F$-extremization refers to extremizing the $S^{3}$ free energy with respect to the trial $R$-charges to obtain the superconformal fixed point \cite{Jafferis:2010un}. 

The gravity duals to the four-dimensional theories are given by the backreacted geometry of a stack of $N$ D3-branes located at a CY$_{3}$ singularity. The near-horizon geometry is given by AdS$_{5}\times\text{SE}_{5}$ in type IIB supergravity, where $\text{SE}_{5}$ is the base of the CY$_{3}$ cone. On the gravity side the $a$ central charge is related to the volume of this SE$_{5}$ manifold through \cite{Henningson:1998gx}
\begin{gather}\label{a-function}
	a=\frac{\pi^{3}}{4\text{vol}(\text{SE}_{5})}N^{2}\,.
\end{gather}
Due to the intimate relation between the parent and daughter theories, this function will appear in the expressions of the observables of the three-dimensional theories that we list below.

The way in which the parent theories are related to the three-dimensional $\mathcal{N}=2$ SCFTs of \famb can be made more transparent by looking at the corresponding gravitational duals. As briefly mentioned in the introduction, the holographic dual to the Chern-Simons-matter theories of this family, is given by a solution in massive type IIA supergravity. In practice, one can arrive at the latter solution, upon T-dualizing the type IIB solution, which is dual to the higher-dimensional parent theory. 
By turning on a Romans mass $F_{0}$, which corresponds to a non-trivial sum over the Chern-Simons levels $k_{i}$, one finds the Chern-Simons-matter theory that resides on the worldvolume of a stack of $N$ D2-branes. These branes probe the more intricate geometry of $\mathbf{R}\times \mathcal{S}(\text{SE}_{5})$, where $\mathcal{S}(\text{SE}_{5})$ denotes the sine-cone over the five dimensional Sasaki-Einstein manifold, whose metric reads
\begin{gather}\label{sine cone metric}
	\dd{s}^{2}_{\mathcal{S}(\text{SE}_{5})}=\dd{\al}^{2}+\sin^{2}\al\dd{s}^{2}_{\text{SE}_{5}}\,.
\end{gather} 
The resulting near-horizon geometry is topologically a warped product of AdS$_{4}$ and a deformed sine-cone over SE$_{5}$ in massive type IIA supergravity. 

In light of the relation between the parent and daughter theories, the sphere free energy can be expressed in terms of the $a$-anomaly coefficient of the parent theory \cite{Fluder:2015eoa}
\begin{gather}\label{Free energy fam B}
	\text{Re}F(R)=\frac{2^{5/3}3^{1/6}\pi^{3}}{5}a(R)^{2/3}(nN)^{1/3}\,,
\end{gather}
with $n=2\pi\ell_{s} F_{0}$.
In the same paper, a similar result for the vacuum expectation value of the Wilson loop in the large $N$ limit was obtained
\begin{gather}\label{WL fam B large N}
	\text{Re}\log\langle W\rangle=\frac{2^{4/3}\pi}{3^{1/6}}\qty(\frac{a}{nN})^{1/3}\,,
\end{gather}
which matches the partition function of a fundamental string at leading order. This universal expression in the large $N$ limit motivates us to explore whether one can extract a similarly universal expression beyond leading order. This is the aim of this section, namely to analyse the holographic Wilson loop for Chern-Simons-matter theories of \famb beyond the large $N$ and strong coupling limit. We achieve this by studying the partition function of a fundamental string in the massive type IIA string theory background at one-loop order, which allows us to make a prediction for the subleading correction to the vev of the Wilson loop.

Let us start our analysis by presenting the solution of the holographic background of these SCFTs.\footnote{A more general class of AdS$_{4}$ backgrounds was identified in \cite{Passias:2018zlm}.} This solution was first introduced in \cite{Guarino:2015jca} for a specific Chern-Simons-matter theory of this family, namely a three-dimensional SCFT with gauge group $SU(N)$ and Chern-Simons level $k$. In \cite{Fluder:2015eoa} this solution was further generalized to a class of SCFTs in family B. The holographic dual solutions are given by the following warped AdS$_{4}\times X_{6}$ type IIA supergravity backgrounds with $X_{6}\cong \mathcal{S}(\text{SE}_{5})$, whose string frame metrics read
\begin{equation}\label{general metric string frame}
	\dd{s}^{2}=L^{2}\e^{\phi_{0}/2}\sqrt{5+\cos 2\al}\Big[\dd{s_{\text{AdS}_{4}}^{2}}+\frac{3}{2}\dd{\al}^{2}+\frac{6\sin^{2}\al}{3+\cos 2\al}\dd{s_{\text{KE}_{4}}^{2}}+\frac{9\sin^{2}\al}{5+\cos 2\al}\eta^{2}\Big]\,,
\end{equation}
with the dilaton and Romans mass respectively given by
\begin{equation}
	\e^{\Phi}=\e^{\phi_{0}}\frac{(5+\cos 2\al)^{3/4}}{3+\cos 2\al}\,,\qquad F_{0}=\frac{\e^{-5\phi_{0}/4}}{\sqrt{3}L}\,,
\end{equation}
and the following fluxes turned on
\begin{equation}\label{fluxes mIIA}
	\begin{aligned}
		H_{3}&=\dd B_{2}=24\sqrt{2}L^{2}\e^{\phi_{0}/2}\frac{\sin^{3}\al}{(3+\cos2\al)^{2}}J\wedge\dd{\al}\,,\\
		F_{2}&=-L \e^{-3\phi_{0}/4}\sqrt{6}\Big(4\frac{\sin^{2}\al\cos\al}{(3+\cos2\al)(5+\cos2\al)}J \\ &\qquad\qquad\qquad\qquad+3\frac{3-\cos2\al}{(5+\cos2\al)^{2}}\sin\al\dd{\al}\wedge\eta\Big)\,,\\
		F_{4}&=L^{3} \e^{-\phi_{0}/4}\Big(6\,\text{vol}_{\text{AdS}_{4}}+12\sqrt{3}\frac{7+3\cos2\al}{(3+\cos2\al)^{2}}\sin^{4}\al\,\text{vol}_{\text{KE}_{4}}\\ &\quad\quad\quad\quad\quad+18\sqrt{3}\frac{(9+\cos2\al)\sin^{3}\al\cos\al}{(3+\cos2\al)(5+\cos2\al)}J\wedge\dd{\al}\wedge\eta\Big)\,,
	\end{aligned}
\end{equation}
where the range of the angle $\al$ is $\al\in[0,\pi]$. Moreover, $\dd{s^{2}_{\text{KE}_{4}}}$ denotes the metric of a four-dimensional Kähler-Einstein manifold and $\eta$ is its associated contact one-form. Together these comprise a five-dimensional Sasaki-Einstein manifold whose metric is given by
\begin{gather}
	\dd{s}_{\text{SE}_{5}}^{2}=\dd{s}^{2}_{\text{KE}_{4}}+\eta^{2}\,.
\end{gather}
It is easily observed that the metric on $X_{6}$ is topologically equivalent to the sine-cone over $\text{SE}_{5}$. Moreover, the flux quantization condition in this case yields \cite{Fluder:2015eoa}
\begin{gather}\label{flux quant fam B} 		
	\qty(\frac{L}{\ell_{s}})^{6}=\frac{\pi^{6} n^{1/4}}{2^{7/8}3^{7/4}}\qty(\frac{N}{\text{vol}(\text{SE}_{5})})^{5/4}\,,\qquad \e^{\phi_{0}}=\frac{2^{11/12}}{3^{1/6}n^{5/6}}\qty(\frac{\text{vol}(\text{SE}_{5})}{N})^{1/6}\,.
\end{gather} 
\subsection{String dual to Wilson loop}

Now, let us turn to the evaluation of the string partition function, which in the large $N$ limit is holographically dual to the vacuum expectation value of a $1/2-$BPS Wilson loop in the fundamental representation. The relevant partition function is that of a fundamental string in the AdS$_{4}$ background, whose worldsheet is given by the minimal surface of $\Sigma_{2}\subset\text{AdS}_{4}$, this reduces to the following holographic relation
\begin{gather}
	\langle W\rangle=Z_{\text{string}}\approx \e^{-S_\text{cl}}Z_{\text{1-loop}}\,,
\end{gather}  
where we expanded the string partition function for strong 't~Hooft coupling $\tilde\la=N/n$ around the classical extremal surface.
The classical string action in this case reduces to the regularized area of the string worldsheet and was matched with the leading order $\mathcal{O}(N^{1/3})$ of the Wilson loop in \cite{Fluder:2015eoa}. For completeness, we will repeat this calculation below. We furthermore evaluate the one-loop partition function, which allows us to make a prediction for the prefactor of the exponent appearing in the Wilson loop vacuum expectation value.

In the Green-Schwarz formalism, the string action consists of three terms. First is the familiar Polyakov action,
\begin{gather}\label{string cl action}
	S_{\text{string}}^{\text{bos}}=\frac{1}{4\pi\ell_{s}^{2}}\int\gamma^{ij}G_{ij}\text{vol}_{2}+\frac{i}{2\pi\ell_{s}^{2}}\int P\left[B_{2}\right]\,,
\end{gather}
where we denote the target-space metric with $G_{\mu\nu}$ and its pull-back to the worldsheet with $G_{ij}=G_{\mu\nu}\partial_{i}X^{\mu}\partial_{j}X^{\nu}$ with the worldsheet coordinates denoted by the latin indices $i,j=1,2$ and $\mu,\nu=1,\dots,10$. The Polyakov action encodes the dynamics of the scalar degrees of freedom on the string and their coupling to the background Kalb-Ramond field whose pull-back to the worldsheet is denoted by $P\left[B_{2}\right]$. In addition, we also have couplings to the worldsheet fermions, as well as a coupling to the background dilaton both of which we will return to momentarily. 
Note that all worldsheet and background quantities that appear in this section are two- and ten-dimensional respectively. These should not be confused with the three- and eleven-dimensional quantities we encountered in Section \ref{sec:famA gravity duals}. 

For the case at hand we identify the classical string configuration by finding the relevant extremal surface that is dual to the 1/2-BPS Wilson loop. To this end, as for the M2-brane before, we work in static gauge where we have $\partial_{i}X^{\mu}=\delta_{i}^{\mu}$, which implies $G_{ij}=\gamma_{ij}$. The boundary conditions for the string together with the equation of motion essentially fixes the string configuration to an AdS$_{2}\subset\text{AdS}_{4}$ that is located around $\al=0$.
This results in the following expression for the induced worldsheet metric
\begin{gather}\label{string ws metric}
	\dd{s}_{\text{string}}^{2}=\frac{\ell^2}{\sinh^{2}\s}(\dd{\s}^{2}+\dd{\tau}^{2})\,,\qquad \ell^2 = \sqrt{6}L^{2}e^{\phi_{0}/2}\,,
\end{gather}
where we have defined the effective AdS$_2$ length scale $\ell$.
Using this, together with the fact that the $B$-field does not contribute, as it has no legs on the worldsheet, the classical string action is given by \cite{Fluder:2015eoa}
\begin{gather}\label{class action fam B}
	S_{\text{cl}}=-\pi^{2}\qty(\frac{4N}{\sqrt{3}n\,\text{vol}(\text{SE}_{5})})^{1/3}\,,
\end{gather}
which matches \eqref{WL fam B large N}. To arrive at this result, we used the regularized expression vol$(\text{AdS}_{2})=-2\pi$ together with the flux quantization conditions \eqref{flux quant fam B}.

\subsubsection{The quadratic action}

We continue our analysis by studying the one-loop string partition function. Expanding the Polyakov action to quadratic order in static gauge results in a Gaussian field theory for eight scalar fields living on the classical worldsheet \eqref{string ws metric}. Much like for the M2-brane, the quadratic action for the scalars can be derived in generality and the result can be expressed in terms of the background geometric quantities. In an effort to keep this section concise we will refrain from giving too many details and only give the final result \cite{Singh:2023olv} 
\be\label{one-loop bos lagrangian}
\begin{split}
	S^{(2)}_{\text{bos}}&=\frac{1}{2\pi\ell_{s}^{2}}\int\text{vol}_{2}\mathcal{L}_{\text{bos}}\,,\\
	\mathcal{L}_{\text{bos}}&=\frac{1}{2}(\tensor{D}{^{a}_{b}}\zeta^{b})^{2}-\frac{1}{2}\Big(\tensor{R}{^{i}_{aib}}+\tensor{K}{_{a}^{ij}}\tensor{K}{_{bij}}-\frac{i}{2}\epsilon^{ij}\nabla_{a}H_{bij}-\frac{1}{4}\tensor{H}{_{a}^{ci}}\tensor{H}{_{bci}}\Big)\zeta^{a}\zeta^{b}\,,
\end{split}\ee
where the covariant derivative in this case reads
\begin{gather}
	\tensor{D}{^{a}_{b}}\zeta^{b}=\nabla\zeta^{a}+ \tensor{A}{^{a}_{b}}\zeta^{b}\,,\qquad \tensor{A}{_{i}^{ab}}=\tensor{\Omega}{_{i}^{ab}}+\frac{i}{2}\epsilon_{ij}\tensor{H}{^{jab}}\,.
\end{gather}
Again we denote the fluctuations normal to the worldsheet with $\zeta^{a}$ and $a,b=1,\dots,8$, while $i,j=1,2$ are the curved worldsheet indices. The three-form is given by $H_{3}=\dd{B_{2}}$. Leaving the details for Appendix \ref{app:spectrum fam B}, this rather unwieldy expression reduces to a collection of Klein-Gordon actions for scalars on AdS$_2$. As suggested by the symmetry of the underlying worldsheet geometry, the connection vanishes and the only non-trivial data that remains are the masses of the fields. These are computed in Appendix \ref{app:spectrum fam B} and summarized in Table \ref{tab: 2d spectrum}.
\begin{table}[h!]
	\begin{center}
		{\renewcommand{\arraystretch}{1.3}
			\begin{tabular}{@{\extracolsep{10 pt}}llc}
				\toprule
				Field & d.o.f. & $M^2\ell^2$  \\\midrule
				Scalars & $2$ & $2$  \\
				&$6$  & $-2/9$ \\\midrule
				Fermions & 	$2$ & $1$   \\
				& 	$6$ & $4/9$ \\
				\bottomrule
		\end{tabular}}
		\caption{Spectrum of bosonic and fermionic string fluctuations.}
		\label{tab: 2d spectrum}
	\end{center}
\end{table}

In addition to the scalar fields, we also have the fermionic Green-Schwarz action
\cite{Cvetic:1999zs}
\begin{equation}
	\begin{aligned}
		S_{\text{ferm}}=-\frac{i}{2\pi\ell_{s}^{2}}\int\bar\theta \mathcal{P}^{ij}\Big\{\Gamma_{i}D_{j}-&\frac{1}{8}\Gamma_{11}\Gamma_{i}^{\mu\nu}H_{j\mu\nu}+\frac{1}{8}F_{0} \e^{\Phi}\Gamma_{i}\Gamma_{j}\\
		&-\frac{1}{8}\e^{\Phi}\qty(\Gamma_{11}\Gamma_{i}\slashed{F}_{2}\Gamma_{j}-\Gamma_{i}\slashed{F}_{4}\Gamma_{j})\Big\}\theta\dd{\s}\dd{\tau}\,,
	\end{aligned}
\end{equation} 
with $\mathcal{P}^{ij}=\sqrt{\g}\g^{ij}-i\Gamma_{11}\epsilon^{ij}$. Here $\epsilon^{ij}$ is the Levi-Civita symbol, $\theta$ is a 32-component Dirac spinor and $\Gamma_{i}$ are the ten-dimensional gamma matrices. As for the bosonic action, this reduces to a collection of standard fermionic actions on AdS$_2$ where the only non-trivial information inherited by the background are the fermionic masses. Once again we refer to Appendix \ref{app:spectrum fam B} for the explicit computation, but the result is summarized in Table \ref{tab: 2d spectrum}.

Finally, we must take into account the coupling of the worldsheet to the dilaton, which is captured by the Fradkin-Tseytlin action. The latter is given in terms of the pull-back of the dilaton to the worldsheet, which we denote as $\Phi_{0}$, and reads
\begin{gather}\label{FT action}
	S_{\text{FT}}=\frac{1}{4\pi}\int\text{vol}_{2}\Phi_{0} R^{(2)}+\frac{1}{2\pi}\int_{\partial}\dd{s} \Phi_{0} K\,,
\end{gather}
where the Ricci scalar of the worldsheet is given by $R^{(2)}$ and $K$ denotes the extrinsic curvature scalar on the boundary of the worldsheet. This classical contribution must be combined with the previous terms coming from the quantum fluctuations. This is because already at classical level it scales as ${\cal O}(\ell/\ell_s)^0$, i.e. in the same way as the contribution of the fluctuations.
In the case at hand, the pull-back of the dilaton is constant
\be
\e^{\Phi_0}=\frac{6^{3/4}}{4}\e^{\phi_{0}}\,.
\ee
As a result, the FT action reduces to $S_{\text{FT}}=\chi\Phi_{0}$, where the Euler characteristic of the string worldsheet is $\chi=1$.  
Notice that none of the terms in the quadratic action or the coupling to the dilaton are sensitive to which KE$_4$ space appears in the ten-dimensional solution \eqref{general metric string frame} and as such the result so far is completely universal. Surely, the relation between supergravity and field theory parameters in \eqref{flux quant fam B} involves the volume of the SE$_5$, which will appear in the final result.

Let us now turn to evaluating the one-loop string partition function or equivalently its quantum effective action which is defined as 
\begin{gather}\label{string eff action general}
	Z_\text{1-loop} \equiv \e^{-\Gamma_{\text{string}}}=\e^{-\Phi_{0}}\int\left[D\zeta D\theta D\bar\theta\right]e^{-S^{(2)}_\text{string}}\,.
\end{gather}
For convenience we integrate by parts and express the fluctuation action as a sum of quadratic operators 
\begin{gather}\label{string quad action }
	S^{(2)}_{\text{string}}=\frac{1}{4\pi \ell_{s}^{2}}\int\text{vol}_{2}(\zeta^{a}\mathcal{K}_{ab}\zeta^{b}+\bar\theta^{a}\mathcal{D}_{ab}\theta^{b})\,,
\end{gather}
where we denote the quadratic scalar and fermionic operators by $\mathcal{K}_{ab}$ and $\mathcal{D}_{ab}$ respectively. The operators are in fact diagonal in the $a,b$-indices and as discussed are only characterized by the masses of the fields
\begin{gather}
	\mathcal{K}=-\nabla^{2}+M^{2}\,,\qquad \mathcal{D}=i\slashed{\nabla}+a\s_{3}+v\,,
\end{gather}
where for the fermions $M^2 = a^{2}-v^{2}$ is the mass of the appropriately squared fermionic operator. This allows us to perform the path integral in \eqref{string eff action general} and reduce it to a product of functional determinants of the kinetic operators
\begin{gather}\label{prod of dets}
	\Gamma_{\text{string}}=\Phi_0 + \frac{1}{2}\log\frac{\displaystyle \prod_{\text{bosons}}\det \mathcal{K}}{\displaystyle \prod_{\text{fermions}}\det \mathcal{D}}\,.
\end{gather}

At this point we note that remarkably our spectrum of fluctuations in Table \ref{tab: 2d spectrum} matches the string spectrum we encountered when studying the holographic dual to Wilson loops in the $V^{5,2}$ model in section \ref{sub:v52}. Recall that in this case we had $q_{l}=\{3,-1/3,-1/3,-1/3\}$. The corresponding string masses can be read off from Table \ref{tab: 3d spectrum} together with \eqref{reductionto2D} for $n=0$. The result precisely matches the entries in Table \ref{tab: 2d spectrum}. This means that we can essentially borrow the results from Section \ref{sub:v52} when computing the operator determinants, resulting in
\be
\Gamma_{\text{string}} = \Phi_{0}+\log (2\pi)-3\log\Gamma\Big(\f23\Big) -\log\Lambda\,.
\ee
A crucial difference here is that in isolation, the string partition function (corresponding to level $n=0$ in section \ref{sub:v52}) is logarithmically divergent. When combined with the infinite tower of Fourier modes on the M2-brane this logarithmic divergence cancels \cite{Giombi:2023vzu}, but in the present case, for the string we must regularize it with a UV cutoff $\Lambda$. This log-divergence is universally encountered in the string partition function computation and is by now well understood. For instance, taking a ratio of two one-loop string partition functions whose string worldsheets have the same topology leads to a finite result. Ultimately this amounts to replacing $\log\Lambda$ by $\log(\ell/(2\pi\ell_s))$. We refer to \cite{Drukker:2000ep,Giombi:2020mhz,Gautason:2021vfc,Astesiano:2024sgi} and references therein for a more complete discussion. Performing this replacement, and after some algebra, the final result reads
\be
\e^{-\Gamma_{\text{string}}} = \Gamma\qty(\frac{2}{3})^{3}\f{L}{\sqrt{6}\pi^2 \ell_s}\e^{-3\phi_0/4}=\Gamma\qty(\frac{2}{3})^{3}\frac{n^{2/3}N^{1/3}}{2^{4/3}3^{2/3}\pi\,\text{vol}(\text{SE}_{5})^{1/3}}\,.
\ee

Combining this answer with the classical string area, we are now able to use the holographic duality to obtain the vacuum expectation value of the $1/2$-BPS Wilson loop of \famb up to next-to-leading order
\begin{gather}\label{WL fam B}
	\langle W\rangle_{\text{B}}\approx  \Gamma\qty(\frac{2}{3})^{3}\frac{n^{2/3}N^{1/3}}{2^{4/3}3^{2/3}\pi\,\text{vol}(\text{SE}_{5})^{1/3}}\exp\qty(\pi^{2}\frac{4^{1/3}N^{1/3}}{3^{1/6}n^{1/3}\,\text{vol}(\text{SE}_{5})^{1/3}})\,.
\end{gather}
%\ffg{here}
Applying this expression to the cases where the SE$_5$ manifold is $S^{5}$, $T^{1,1}$ \cite{Romans:1984an} and the infinite class $Y^{p,q}$ \cite{Gauntlett:2004yd}, yields a novel result for the respective holographic Wilson loop vev up to next-to-leading order.
Naturally, we can express this result in terms of the 't Hooft coupling constant $\tilde{\la}=N/n$, which is altered accordingly to the case of $\sum_{i=1}^{\mathcal{G}}k_{i}=n$ for the SCFTs of family B. In this case, we can rewrite \eqref{WL fam B} as
\begin{gather}\label{eq:WL nlo fam B}
	\langle W\rangle_{\text{B}}\approx  \Gamma\qty(\frac{2}{3})^{3}\frac{N}{2^{4/3}3^{2/3}\pi\,\text{vol}(\text{SE}_{5})^{1/3}\tilde{\la}^{2/3}}\exp\qty(\pi^{2}\frac{4^{1/3}\tilde{\la}^{1/3}}{3^{1/6}\,\text{vol}(\text{SE}_{5})^{1/3}})\,.
\end{gather}
Since the subleading correction to the WL of family B has not been studied before in the literature and with no field theory answer to compare to, one might propose that there is no evidence in support of the above prediction. However, by comparing the scaling of \eqref{eq:WL nlo fam B} with respect to $\tilde\lambda$ and $N$, with the respective scaling of the WL of family A, one can show that the above result has indeed the expected scaling. One can also gain more intuition by comparing how the free energy of family A and B scale with the respective field theory parameters. As reviewed in Section \ref{sec:2SCFTfams} the free energy on $S^{3}$ of family A and B respectively, behaves in the large $N$ limit in the following manner
\begin{gather}
	F_{\text{A}}\sim\frac{N^{2}}{\sqrt{\la}}\,,\qquad F_{\text{B}}\sim \frac{N^{2}}{\tilde{\la}^{1/3}}\,.
\end{gather}
From here we observe that the scaling with the rank of the gauge group $N$ is the same for both families, while the power dependence on the corresponding 't Hooft coupling changes from $\la^{-1/2}$ in family A to $\tilde\la^{-1/3}$ for family B.\footnote{In fact, the authors of \cite{Hong:2021bsb} provided evidence that the subleading correction in the 't Hooft coupling in the large $N$ limit of the $S^{3}$ free energy of a specific Gaiotto-Tomasiello SCFT that also has a dual description in massive type IIA supergravity, scales as
	\begin{gather}
		F\sim N^{2}(\frac{\mathfrak{c}_{1}}{\tilde\la^{1/3}}+\frac{\mathfrak{c}_{2}}{\tilde\la}+\dots)\,,
	\end{gather}
	where $\mathfrak{c}_{1},\mathfrak{c}_{2}$ are constants that can be found in \cite{Hong:2021bsb}. This is in agreement with the analogous behaviour that has been found for the free energy of the SCFTs of family A which behave as 
	\begin{gather}
		F_{\text{A}}\sim N^{2}(\frac{\mathfrak{a}_{1}}{\la^{1/2}}+\frac{\mathfrak{a}_{2}}{\la^{3/2}}+\dots)\,,
	\end{gather}
	where $\mathfrak{a}_{1},\mathfrak{a}_{2}$ are again constants fixed by the argument of the Airy function.} Motivated by this, we expect that if the next-to-leading order (NLO) contribution to the WL of family B scales as  
\begin{gather}
	\log \langle W\rangle_{\text{B}}^{\text{NLO}}\sim \log\frac{N}{\tilde{\la}^{2/3}}\,,
\end{gather}
then the NLO correction to the WL of family A should scale as
\begin{gather}\label{eq:scaling WL A}
	\log\langle W\rangle_{\text{A}}^{\text{NLO}}\sim\log\frac{N}{\la}\,.
\end{gather}
This is exactly how the NLO piece of the WL of family A scales with its corresponding field theory parameters in the  large $N,k$ limit, with $\lambda$ fixed
\begin{gather}
	\langle W\rangle_{\text{A}}\approx  \frac{N}{4\pi c\la}\exp\qty(2\pi^{3}c \sqrt{\frac{\la}{6\, \text{vol(SE$_{7}$)}}})\,.
\end{gather}
Therefore, we expect that at least the scaling of \eqref{eq:WL nlo fam B} with respect to the corresponding field theory parameters $N, \tilde\la$ is in agreement  with what one would obtain for the NLO correction to the WL by studying the matrix model of family B. We hope to report some progress on this topic in the future.

%----------------------------------------------------------------------------------------
%	Summary
%----------------------------------------------------------------------------------------
\section{Conclusions and future prospects}
\label{sec:summary}
%%%%%%%%%%%%%%%%%%%%%%%%%%%%%%%%

In this paper, we provide a novel prediction for the subleading behaviour of the 1/2-$\rm{BPS}$ WL for three-dimensional $\mathcal{N}=2$ Chern-Simons-matter theories belonging to two distinct families.  We achieve this by using the holographic description of these SCFTs. More concretely, we find two universal expressions for the subleading correction to the holographic WLs of family A and B. The former is given by the one-loop partition function of a probe M2-brane, while the latter is dual to the one-loop partition function of a fundamental string. Irrespective of the dual eleven- or ten-dimensional geometry, the worldvolume of the M2-brane is that of AdS$_{2}\times S^{1}_{M}$, with $S^{1}_{M}\subset \rm{SE}_{7}$ being an appropriately calibrated M-theory circle \cite{Farquet:2013cwa}, while the minimal surface of the string corresponds to that of AdS$_{2}$. We find that the underlying eleven- and ten-dimensional geometries allow for a universal analysis of the semiclassical quantization of the membrane and string, respectively. As a result, the expression for the one-loop M2-brane partition function can be read off from (\ref{eq:gamma with qs} - \ref{M2 part funct}), which is a function of the on-shell radius of the M-theory circle $c$ that can be directly extracted from the metric of the internal seven-dimensional geometry and the charges $q_{l}$, which can be determined from the relevant spin connection on the SE$_{7}$ manifold. On the other hand, the one-loop partition function of the string in the massive type IIA supergravity backgrounds is given by \eqref{WL fam B} and only depends on the number of D2-branes $N$, the Romans mass $n$ and the volume of the underlying SE$_{5}$ manifold. These results are valid for an infinite class of geometries and as such provide a prediction for the subleading correction to the WL operator for an infinite class of SCFTs.

We combined our universal prediction for the subleading correction to the WL of family A with the results of  \cite{Gautason:2025plx}, where it was argued that the M2-brane partition function yields the vev of the WL in the grand canonical ensemble (instead of the canonical ensemble). Hence, before comparing our results with the field theory answer we first perform a Laplace transform. This led us to a closed form expression for the perturbative part of the vacuum expectation value of the $1/2-\rm{BPS}$ WL of family A
\begin{gather}\label{eq:WL conjecture}
	\langle W(N)\rangle^{\rm{p}}=\e^{-\Gamma_{\text{M2}}}\frac{\text{Ai}(\mathcal{C}^{-1/3}(N-2c-\mathcal{B}))}{\text{Ai}(\mathcal{C}^{-1/3}(N-\mathcal{B}))}\,,
\end{gather} 
where $\e^{-\Gamma_{\text{M2}}}$ is the one-loop M2-brane partition function, which is given by (\ref{eq:gamma with qs} - \ref{eq: M2 eff action sum}). We highlight that this is merely the $1/N-$perturbative part of the WL operator and does not include the non-perturbative corrections, which would be exponentially suppressed in $N$ and be of $\mathcal{O}(\e^{-\# \sqrt{N}})$. Our findings of Section \ref{sec:examples famA} suggest that the contribution coming from the one-loop M2-brane partition function $\e^{-\Gamma_{\text{M2}}}$  seems to always be given in terms of trigonometric functions, when the internal space is a \textit{toric} Sasaki-Einstein manifold.\footnote{The only example for which $\e^{-\Gamma_{\text{M2}}}$ is not given in terms of elementary functions is for the $V_{5,2}$ manifold, which is non-toric.} Some exceptions worth highlighting are the $Q^{1,1,1}/\mathbf{Z}_{k}$, $Q^{2,2,2}/\mathbf{Z}_{k}$ and $M^{3,2}/\mathbf{Z}_{k}$ models. Surprisingly, for the first we find $\e^{-\Gamma_{\text{M2}}}=k$ while for the latter two models we obtain $\e^{-\Gamma_{\text{M2}}}=2k$, indicating that the string theory limit $k\gg 1$ for each of these theories is exact. 
The results for the $Q^{2,2,2}/\mathbf{Z}_{k}$ and $M^{3,2}/\mathbf{Z}_{k}$ models seem to be in agreement with the fact that the recent analysis performed for the $Q^{1,1,1}$ and $D_{3}$ models in \cite{Hosseini:2025jxb} is not straightforwardly applicable to these former chiral quivers, as they might lack an M-theory phase. From our perspective the M-theory and string theory limits are indistinguishable.
As highlighted earlier, in general the M2-brane partition function is sensitive to the three charges $q_{l}$ that can be read off from the SE$_{7}$ metric. Although their connection to the dual field theory remains unclear, they seem to be related to the $R$-charges of the matter fields.

Motivated by our results, there are a variety of future directions that are worth exploring. First, and most obviously, it would be interesting to verify our results from a matrix model perspective. It is likely that for the gauge theories of family A, this would require employing similar numerical methods as were used in \cite{Geukens:2024zmt,Bobev:2025ltz}. Regarding family B, the authors of \cite{Hong:2021bsb} studied the matrix model of a certain class of Gaiotto-Tomasiello-like theories \cite{Gaiotto:2009yz, Gaiotto:2009mv}, which do not belong to the SCFTs of family B that we study. These theories are $\mathcal{N}=3$ Chern-Simons-matter necklace quivers with $r$-nodes, where $r\geq 2$ and feature a non-trivial Romans mass. It would be interesting to explore whether the methods used in \cite{Hong:2021bsb} could be employed to study the corresponding matrix model of the theories we study. It would also be interesting to perform a similar holographic analysis for the $\mathcal{N}=3$ Gaiotto-Tomasiello-like theories. We plan to discuss our findings on these topics in upcoming work \cite{FuturePaper}.

In an effort to extent our current analysis, it would be interesting to explore whether or not other observables enjoy a universal description similar to the one we recovered for the vacuum expectation value of the holographic WL. Evidence for this was already provided from a matrix model perspective that this indeed seems to be the case for the perturbative part of the free energy on $S^{3}$ for selected SCFTs of family A in \cite{Geukens:2024zmt,Bobev:2025ltz}. However, given our universal analysis of the dual geometries of family A, it would be compelling to explore if we could apply our methodology to study worldsheet instantons in the AdS$_{4}\times\rm{SE}_{7}$ backgrounds. These would correspond to non-perturbative corrections to the free energy. As explained in \cite{Cagnazzo:2009zh, Gautason:2023igo} these worldsheet instantons wrap non-trivial 2-cycles in the internal geometry and therefore it is not entirely clear whether such an analysis could be performed in a universal manner. The M-theory counterpart of the worldsheet instanton is an M2-brane instanton that wraps a non-trivial 3-cycle in the internal geometry. In our language the 3-cycle should consist of the M-theory circle $S^{1}_{M}$ and a non-trivial 2-cycle of the SE$_{7}$ geometry. An example of this was studied in \cite{Beccaria:2023ujc,Gautason:2025per} for the case of the M-theory dual to ABJM, matching the instanton prefactor of the large $N$ leading order non-perturbative contribution to the free energy on $S^{3}$.

%----------------------------------------------------------------------------------------
%	ACKNOWLEDGEMENTS
%----------------------------------------------------------------------------------------
%%%%%%%%%%%%%%%%%%%%%%%%%%%%%%%%%%%%%
\bigskip
\bigskip
\leftline{\bf Acknowledgements}
\smallskip
\noindent We are grateful to Christopher Couzens, Seyed Morteza Hosseini, Anayeli Ramírez, James Sparks, and Jesse van Muiden for fruitful discussions. We especially thank Nikolay Bobev, Pieter-Jan De Smet, Valentina Giangreco M. Puletti, Evangelos Tsolakidis, and Konstantin Zarembo for comments on the manuscript. FFG and AN are supported by the Icelandic Research Fund under grant 228952-053. The work of AN is supported by the Eimskip Fund of the University of Iceland, the Aðalsteinn Kristjánsson Memorial Fund and the COST Action CA22113 through an STSM grant. AN is grateful to the University of Southampton, KU Leuven and the University of Oviedo for the warm hospitality where parts of this work were carried out.

\newpage
%----------------------------------------------------------------------------------------
%	APPENDICES
%----------------------------------------------------------------------------------------
\appendix
\section{Sasaki-Einstein properties}
\label{app:SE}
As shown in the main text, a $(2d+1)$-dimensional Sasaki-Einstein (SE) manifold can be written in terms of its transverse Kähler-Einstein (KE) base and contact one-form $\eta$ as follows
\begin{gather}\label{eq:SE metric2}
	\dd{s}_{\text{SE}_{2d+1}}^{2}=\eta^{2}+\dd{s}_{\text{KE}_{2d}}^{2}\,.
\end{gather}
Here we summarize some properties of SE manifolds that will prove useful in the analysis of the M2-brane partition function in AdS$_{4}\times$SE$_{7}$ of Section \ref{sec:famA gravity duals}. We may write the vielbeine associated to the SE metric locally as
\begin{gather}
	\hat{e}^{0}=\eta=\dd\xi+A\,,\quad \hat{e}^{a}=e^{a}\,, 
\end{gather}
where the one-form $A=A_{a}\dd{x^{a}}$ has no leg along $\xi$ and does not depend on $\xi$. Here, the hat refers to the higher-dimensional SE space and $a,b=1,2,\dots,2d$ are the tangent space indices. Now, one can relate the spin connection $\tensor{\hat\Omega}{^{a}_{b}}=\tensor{\hat\Omega}{_{\mu}^{a}_{b}}\dd{x^{\mu}}$ with the Kähler two-form $\omega=\dd{\eta}/2$, where $\omega=\frac{1}{2} J_{ab}e^{a}\wedge e^{b}$, and $J_{ab}$ is the almost complex structure of the KE base manifold, by using the torsion-free condition 
\begin{gather}
	\begin{aligned}
		\tensor{\hat\Omega}{^{0}_{a}}\wedge e^{a}&=-2\omega\,,\\
		\tensor{\hat\Omega}{^{a}_{0}}\wedge \hat e^{0}&= (\tensor{\Omega}{^{a}_{b}}-\tensor{\hat\Omega}{^{a}_{b}})\wedge e^{b}\,.
	\end{aligned} 
\end{gather}
which reduces to
\begin{gather}\label{spin conn and Kahler form}
	\begin{aligned}
		\tensor{\hat\Omega}{^{0}_{b}}&=-J_{ab}e^{a}\,,\\
		\tensor{\hat\Omega}{^{ab}}&=\tensor{\Omega}{^{ab}}-J^{ab}\hat e^{0}\,.
	\end{aligned} 
\end{gather}
After massaging these relations, we are able to deduce the following expressions for the Riemann tensor
\begin{gather}
	\begin{aligned}
		\hat{R}_{abcd}&=R_{abcd}-J_{ac}J_{bd}+J_{ad}J_{bc}-2J_{ab}J_{cd}\\
		\hat{R}_{0a0b}&=J_{ac}J_{bc}\,,\qquad \hat{R}_{abc0}=\nabla_{c}J_{ab}\,.
	\end{aligned}
\end{gather}
Moreover, using the properties of the almost complex structure (i.e. $J_{ab}J_{bc}=-\delta_{ac}$) we find that 
\begin{gather}\label{Riemann SE}
	\hat{R}_{0a0b}=\delta_{ab}\,,
\end{gather}
Finally, using this result in \eqref{quad action M2}, crucially leads to a universal expression for the three-dimensional mass spectrum of the fluctuations about the classical solution of the M2-brane, which is given in Table \ref{tab: 3d spectrum}.

\section{Heat Kernel details}
\label{app:HK}
In the main text of this work we evaluated the functional determinants of the scalar and fermionic quadratic operators by formulating them in terms of the heat kernels of the respective scalar and fermionic Laplacians on AdS$_{2}$ with unit radius. The aim of this section is to provide some of the details related to this well-known analysis, following \cite{Drukker:2000ep} and references therein. As noted previously, we use $\zeta$-regularization to define the functional determinants in terms of their finite and logarithmically UV divergent part. This is achieved by defining the $\zeta$-function of a given quadratic operator $\mathcal{O}=-\nabla^{2}+m^{2}$ in terms of the corresponding trace of the Heat kernel $K(t;m^{2})$ given by
\begin{gather}\label{trace heat kernel}
	K(t;m^{2})=\frac{1}{2\pi^{2}}\int_{0}^{\infty}\dd{\nu}\mu(\nu)\e^{-\la(\nu)t}\,,
\end{gather}
in the following way
\begin{gather}\label{zeta of operator def}
	\zeta(s;m^{2})=-\frac{1}{\Gamma(s)}\int_{0}^{\infty}\dd{t}t^{s-1}K(t;m^{2})=- \frac{1}{2\pi^{2}}\int_{0}^{\infty}\dd{\nu}\mu(\nu)\la(\nu)^{-s}\,,
\end{gather}
where the latter equality is obtained by using the integral representation of $\Gamma(s)$. Here we parametrized the continuous spectrum of eigenvalues with $\nu$ and denoted the eigenvalue function of $\mathcal{O}$ with $\la(\nu)$. In \eqref{trace heat kernel} we also introduced the density of eigenvalues $\mu(\nu)$, hence for the operators under consideration we have
\begin{gather}\label{eigenvalue funct+dens}
	\la_{b/f}(\nu)=\nu^{2}+M^{2}_{b/f}\,,\\
	\mu_{b}(\nu)=\nu\pi\tanh(\nu\pi)\,,\qquad \mu_{f}(\nu)=\nu\pi\cosh(\nu\pi)\,, 
\end{gather} 
with $M_{b}^{2}=m_{b}^{2}+1/4$ and $M^{2}_{f}=m_{f}^{2}$ in the case of scalar and fermionic operators respectively.
In turn, these expressions relate to the functional determinants through the expression
\begin{gather}\label{det with heat kernel}
	\frac{1}{2}\log\det\mathcal{O}=-\frac{1}{2}\text{vol}(\text{AdS}_{2})\int_{\Lambda^{-2}}^{\infty}\dd{t}\frac{K(t;m^{2})}{t}\,,
\end{gather} 
where $\Lambda$ is the UV cut-off scale corresponding to small $t$, which is the limit where \eqref{det with heat kernel} exhibits a logarithmic divergence. It follows from equation \eqref{zeta of operator def} that the divergence is proportional to $\zeta(0;m^{2})$, which corresponds to the Seeley–DeWitt coefficient $a_{2}$. This coefficient can be extracted from the small $t$ expansion of equation \eqref{trace heat kernel} and is given by
\begin{gather}
	a_{2}(\mathcal{O})=\frac{1}{4\pi}\int\text{vol}_{2}b_{2}(\mathcal{O})\,,
\end{gather}
where  we ignored boundary contributions. Here $\text{vol}_{2}$ is the volume form of AdS$_{2}$ (with unit radius) and $b_{2}(\mathcal{O})$ is the local Seeley-DeWitt coefficient, which in our notation reads
\begin{gather}
	b_{2}(\mathcal{K})=\frac{R^{(2)}}{6}-m^{2}_{b}\,,\qquad b_{2}(\mathcal{D}^{2})=-\frac{R^{(2)}}{12}-m^{2}_{f}\,.
\end{gather} 
Combining all the above, we find that 
\begin{gather}
	\frac{1}{2}\log\det\mathcal{O}=-\frac{1}{2}\zeta'(0;m)-\zeta(0;m)\log\Lambda\,,
\end{gather}
where the prime denotes the derivative with respect to $s$ and $\zeta'(0;m)$ can be evaluated explicitly for each operator,
resulting in the following expressions\footnote{At this point, we should note that we defined the density of eigenvalues of the fermionic operator with a relative minus sign compared to \cite{Giombi:2020mhz}. Hence, the zeta function of the fermionic operator derived below is related to the one in \cite{Giombi:2020mhz} through $\zeta'(0;m_{f})_{\text{here}}=-\zeta'(0;m_{f})_{\text{\cite{Giombi:2020mhz}}}$.}
\begin{gather}\label{eq: zeta fncts HK}
	\begin{aligned}
		\zeta'(0;m_{b})&=-\frac{1}{12}(1+\log 2)+\log A-\int_{0}^{m_{b}^{2}+1/4}\dd{x}\psi\Big(\sqrt{x}+\frac{1}{2}\Big)\,,\\
		\zeta'(0;m_{f})&=\frac{1}{6}-2\log A-\abs{m_{f}}-\int_{0}^{m_{f}^{2}}\dd{x}\psi(\sqrt{x})\,,
	\end{aligned}
\end{gather}
where $A$ the Glaisher constant and $\psi(x)=\frac{1}{\Gamma(x)}\derivative{\Gamma(x)}{x}$ the digamma function. Applying our results to the effective action we aim to compute, yields
\begin{gather}\label{eff action HK}
	\begin{aligned}
		\Gamma%&=\frac{1}{2}\log\frac{\displaystyle \prod_{\text{bosons}}\det \mathcal{K}}{\displaystyle \prod_{\text{fermions}}\det \mathcal{D}^{2}}\\
		&=-\frac{1}{2}\Big(\displaystyle \sum_{\text{bosons}}\zeta'(0;m_{b})-\displaystyle \sum_{\text{fermions}}\zeta'(0;m_{f})\Big)-\frac{\log\Lambda}{4\pi}\int\text{vol}_{2}b_{2,\text{tot}}\,,
	\end{aligned}
\end{gather}
with
\begin{gather}
	b_{2,\text{tot}}=\displaystyle \sum_{\tiny{\text{bosons}}}\hspace{-0.15cm}b_{2}(\mathcal{K})-\displaystyle \sum_{\tiny{\text{fermions}}}\hspace{-0.22cm}b_{2}(\mathcal{D}^{2})\,.
\end{gather}
For both the effective action of the string studied in Section \ref{sec:famB gravity duals} and the effective action containing the two-dimensional operators in the case of the M2-brane, 
the total UV divergence is controlled by the Euler characteristic of the worldsheet, as expected. That is, if we take $m^{2}$ to denote the mass-matrix, we have that
\begin{gather}
	b_{2,\text{tot}}=8\frac{R^{(2)}}{6}-\Tr m_{b}^{2}+8\frac{R^{(2)}}{12}+\Tr m_{f}^{2}=R^{(2)}\,,
\end{gather}
and thus the prefactor of $\log\Lambda$ is $\zeta_{\text{tot}}(0)=\chi=1$. In the case of the M2-brane, the effective action \eqref{eq: M2 eff action sum} obviously does not suffer from any divergences, since $\zeta$-regularization yields $\sum_{n=-\infty}^{\infty}1=0$. However, for the case of the effective action of the string, we have to resort to the method of \cite{Giombi:2020mhz} in order to extract its finite answer and make a prediction for the vacuum expectation value of the Wilson loop. The main idea boils down to reinstating the dependence on the AdS$_{2}$ radius $L$, which can be easily done through dimensional analysis. This yields $\Gamma_{\infty}=-\chi\log(L\Lambda)$. Then one should introduce an additional universal counterterm $\log(2\pi\ell_{s}\Lambda)$ that arises from the path integral measure and cancels the UV divergence in \eqref{eff action HK}. 
This regularization procedure is briefly described in Section \ref{sec:famB gravity duals} and we refer to the original paper of \cite{Giombi:2020mhz} for further details.

\section{Universal quadratic operators for gravity duals of family B}
\label{app:spectrum fam B}
In Section \ref{sec:famB gravity duals} we showed that the holographic vacuum expectation value of the Wilson loop for \famb at one-loop order is given by the universal expression of \eqref{WL fam B}, which only depends on the volume of the underlying SE manifold, the Chern-Simons levels whose sum we denote by $n$ and the rank of the gauge group $N$. An essential ingredient of this result is the universal form of the quadratic operators that are acting on the eight scalars and the eight two-dimensional spinor fields. Here we provide some of the details on the derivation of the mass spectrum of these operators. In the first subsection we derive the universal fermionic spectrum and continue with a subsection on the universal scalar mass matrix.
\subsection{Fermionic operators}
Let us start by defining an orthonormal basis for the SE$_{5}$ manifold as follows
\begin{gather}
	\dd{s}_{\text{SE}_{5}}^{2}=\sum_{i=1}^{5}(e^{i})^{2}\,,
\end{gather}
with $e^{i}=\tensor{e}{_{\beta}^{i}}\dd{x}^{\beta}$, where $\beta=1,\dots,5$ and remind the reader of our object of interest, which is to study the fermionic Green-Schwarz (GS) action 
\begin{equation}\nonumber
	S_{\text{ferm}}=-\frac{i}{2\pi\ell_{s}^{2}}\int\bar\theta \mathcal{P}^{ij}\qty{\Gamma_{i}\nabla_{j}-\frac{1}{8}\Gamma_{11}\Gamma_{i}^{\mu\nu}H_{j\mu\nu}-\frac{1}{8}\e^{\Phi}\qty(\Gamma_{11}\Gamma_{i}\slashed{F}_{2}\Gamma_{j}-\Gamma_{i}\slashed{F}_{4}\Gamma_{j})+\frac{1}{8}F_{0} \e^{\Phi}\Gamma_{i}\Gamma_{j}}\theta\dd[2]{\s}\,,
\end{equation} 
with $\mathcal{P}^{ij}=\sqrt{\g}\g^{ij}-i\Gamma_{11}\epsilon^{ij}$ and $\Gamma_{11}=i\Gamma_{\hat\s\hat\tau\hat{x}_{1}\hat{x}_{2}\hat\al\hat{e}_{1}\hat e_{2}\hat e_{3}\hat{e}_{4}\hat e_{5}}$ in our notation. Here $\hat{x}_{1}$, $\hat{x}_{2}$ denote the two normal bundle coordinates of AdS$_{4}$ and the covariant derivative in static gauge reads
\begin{gather}
	\nabla_{i}=\partial_{i}+\frac{1}{4}\Omega^{\hat{\mu}\hat{\nu}}_{i}\Gamma_{\hat{\mu}\hat{\nu}}\,,
\end{gather}
where $\Omega^{\hat{\mu}\hat{\nu}}_{i}$ is the spin connection whose only non-zero component is $\Omega^{\hat{\s}\hat{\tau}}_{\tau}=\coth\s$. 
At this point it is useful to introduce the following notation for the conformal factor of the string worldsheet\footnote{Recall that the worldsheet metric is given by \eqref{string ws metric}.}
\begin{gather}
	\e^{2\rho}=\frac{\ell^{2}}{\sinh^{2}\s}=\frac{4^{1/3}\pi^{2}\tilde{\la}^{1/3}}{3^{1/6}\text{vol}(\text{SE}_{5})^{1/3}\sinh^{2}\s}\,,
\end{gather}
with $\ell^{2}\equiv\sqrt{6}L^{2}\e^{\phi_{0}/2}$.
The fermionic action can be further simplified by observing that the NS-NS field strength $H_{3}=\dd{B_{2}}$ contribution to the action vanishes on-shell, as it has no legs on the string worldsheet. Additionally, we can use that  \cite{Gautason:2021vfc}
\begin{gather}
	\mathcal{P}^{ij}\Gamma_{i}\Gamma_{j}=4 \e^{2\rho}\mathcal{P}\,,\qquad \mathcal{P}^{ij}\Gamma_{i}=2\e^{2\rho}\Gamma^{j}\mathcal{P}\,,
\end{gather}
where we introduced the projector $\mathcal{P}=(\mathbf{1}-i\Gamma_{\hat\s\hat\tau}\Gamma_{11})/2$, to rewrite the GS action in the following way 
\begin{equation}\label{simpler GS action}
	S_{\text{ferm}}=-\frac{i}{\pi\ell_{s}^{2}}\int\bar\theta \qty{\slashed{\nabla}-\frac{1}{4}\e^{\phi}\qty(\Gamma_{11}\slashed{F}_{2}-\tilde{\slashed{F}}_{4})+\frac{1}{4}F_{0} \e^{\phi}}\mathcal{P}\theta\text{vol}_{2}\,.
\end{equation}
We can now fix the $\kappa$-symmetry, that is $\mathcal{P}\theta=\theta$, which reduces the dimensionality of the spinor by half. Note that here we should be careful when permuting the gamma matrices through $\slashed{F}_{4}$ since its first term in \eqref{fluxes mIIA} is proportional to the volume form on AdS$_{4}$ and thus contains legs on $\s$ and $\tau$. Therefore, this specific term acquires a minus sign and we dress the resulting four-form flux with a tilde. To leading order in the classical string solution, the RR fields when contracted with the gamma matrices in tangent space coordinates can be expressed as
\begin{equation}
	\begin{aligned}
		\frac{1}{4}\e^{\Phi}\Gamma_{11}\slashed{F}_{2}&=\frac{i}{24\ell}\Gamma_{\hat\s\hat\tau\hat{\phi}_{1}\hat{\varphi}}\qty(\Gamma_{\hat e_{1}\hat e_{2} \hat e_{3} \hat e_{4}}+ \Gamma_{\hat \al\hat e_{3}\hat e_{4}\hat e_{5}}+\Gamma_{\hat \al\hat e_{1}\hat e_{2}\hat e_{5}})\,,\\
		\frac{1}{4}\e^{\Phi}\tilde{\slashed{F}}_{4}&=\frac{1}{16\ell}\Big(-6i\Gamma_{\hat\s\hat\tau}\Gamma_{\hat{x}_{1}\hat{x}_{2}}+\frac{10}{\sqrt{3}}\qty(\Gamma_{\hat \al\hat e_{1}\hat e_{2}\hat e_{5}}+ \Gamma_{\hat \al\hat e_{3}\hat e_{4}\hat e_{5}}+ \Gamma_{\hat e_{1}\hat e_{2} \hat e_{3} \hat e_{4}})\Big)\,.
	\end{aligned}
\end{equation}
and the Romans mass term reads 
\begin{gather}
	\frac{1}{4}\e^{\Phi}F_{0}=\frac{\sqrt{3}}{8\ell}\,.
\end{gather} 
Plugging these results into \eqref{simpler GS action} yields a 32-dimensional fermionic operator, which can be decomposed to a two-dimensional operator by introducing the following mutually commuting projectors 
\begin{equation}\label{projectors}
	\mathcal{P}_{1,\pm}=\frac{\boldsymbol{1}_{32}\pm i\Gamma_{\hat{x}_{1}\hat x_{2}}}{2}\,,\quad \mathcal{P}_{2,\pm}=\frac{\boldsymbol{1}_{32}\pm \Gamma_{\hat \al\hat e_{1}\hat e_{2}\hat e_{5}}}{2}\,,\quad \mathcal{P}_{3,\pm}=\frac{\boldsymbol{1}_{32}\pm \Gamma_{\hat \al\hat e_{3}\hat e_{4}\hat e_{5}}}{2}\,.
\end{equation}
Here we notice that $\Gamma_{\hat \al\hat e_{3}\hat e_{4}\hat e_{5}}\Gamma_{\hat \al\hat e_{1}\hat e_{2}\hat e_{5}}=-\Gamma_{\hat e_{1}\hat e_{2}\hat e_{3}\hat e_{4}}$, so if $s_{i}=\pm 1$ are the eigenvalues of the projector $\mathcal{P}_{i}$ we find that $\kappa$-symmetry implies $\Gamma_{\hat e_{1}\hat e_{2}\hat e_{3}\hat e_{4}}\theta=-s_{2}s_{3}\theta$. Moreover, using an appropriate representation of the ten-dimensional gamma matrices we identify $\Gamma_{\hat{\s}}=\s_{1}\otimes\mathbf{1}_{16}$ and $\Gamma_{\hat{\tau}}=-\s_{2}\otimes\mathbf{1}_{16}$, where $\s_{m}$, with $m=1,2,3$ are the Pauli matrices. In this case the fermionic action reads
\begin{equation}
	\begin{aligned}
		S_{\text{ferm}}=-&\frac{1}{\pi\ell_{s}^{2}}\int\bar\theta  \Bigg(i\slashed{\nabla}+s_{1}\s_{3}\frac{\mathcal{S}-9}{24}+i\frac{3+5\mathcal{S}}{8\sqrt{3}}\Bigg)\theta\text{vol}_{2}\,,				
	\end{aligned}
\end{equation}
where we set the AdS radius to unity and we defined $\mathcal{S}\equiv s_{2}+s_{3}-s_{2}s_{3}$, which takes on the values $\mathcal{S}=\{1_{\times 6},-3_{\times 2}\}$, where the subscript denotes the respective multiplicity.
Then the two-dimensional fermionic operator is given by
\begin{equation}\label{ferm op IIA}
	\mathcal{D}(s_{1},\mathcal{S})=i\slashed{\nabla}+a(s_{1},\mathcal{S})\s_{3}+v(\mathcal{S})
\end{equation}
with
\begin{equation}
	a(s_{1},\mathcal{S})=s_{1}\frac{\mathcal{S}-9}{24}\,,\quad 
	v(\mathcal{S})=i\frac{3+5\mathcal{S}}{8\sqrt{3}}\,,
\end{equation}
where $v$ has no dependence on the eigenvalue $s_{1}$. From here it is obvious that the fermionic masses are completely independent of the relevant SE manifold under consideration.
\subsection{Bosonic operators}
Let us now turn to the bosonic mass spectrum. This can be derived by studying the one-loop bosonic Lagrangian density given by \eqref{one-loop bos lagrangian}. A first obvious simplification that occurs, is that the relevant NS-NS three-form flux contribution vanishes, as it has no component along any of the worldsheet directions. A second observation we make is that we are studying the fluctuations around the classical configuration of the string where $\al=0$, and the corresponding minimal surface does not extend into the compact part of the target-space metric, this means that the covariant derivatives just reduce to partial derivatives along the worldsheet directions. These observations immediately give rise to the simple expression 
\begin{gather}\label{one-loop bos lagrangian simplified}
	\mathcal{L}_{\text{bos}}=\frac{1}{2}\partial^{i}\zeta_{a}\partial_{i}\zeta^{a}-\frac{1}{2}\Big(\tensor{R}{^{i}_{aib}}+\tensor{K}{_{a}^{ij}}\tensor{K}{_{bij}}\Big)\zeta^{a}\zeta^{b}\,.
\end{gather}
Now, we recall that the extrinsic curvature is given by 
\begin{equation}
	K^{\mu}_{ij}=\partial_{i}\partial_{j}X^{\mu}-\Lambda^{k}_{ij}\partial_{k}X^{\mu}+\Gamma^{\mu}_{\nu \rho}\partial_{i}X^{\nu}\partial_{j}X^{\rho}\,,
\end{equation}
where $\Lambda^{k}_{ij}$ are the Christoffel symbols of the worldsheet metric and $\Gamma^{\mu}_{\nu \rho}$ the Christoffel symbols of the ten-dimensional spacetime metric. Working in static gauge implies freezing the fields which are longitudinal to the worldsheet, and thus $\partial_{i}X^{\mu}=\de^{\mu}_{i}$. It is then straightforward to observe that the extrinsic curvature vanishes at the classical level. Hence, the only non-trivial piece that remains in the mass term of \eqref{one-loop bos lagrangian simplified} is the term proportional to the Riemann tensor. We can then define the so-called bosonic ``mass matrix'' as
\begin{gather}
	\mathcal{M}_{ab}\equiv-\tensor{R}{^{i}_{aib}}\,,
\end{gather} 
and all that remains is to write this Riemann tensor in a universal manner, i.e. show that it is independent of the underlying five-dimensional SE manifold. To achieve this we start by writing the relevant ten-dimensional metric \eqref{general metric string frame} in the following way 
\begin{equation}\label{general warped metric}
	\dd{s}^{2}=\e^{2f(\al)}\dd{s_{\text{AdS}_{4}}^{2}}+\e^{2f(\al)}\Big(\frac{3}{2}\dd{\al}^{2}+\frac{6\sin^{2}\al}{3+\cos 2\al}\dd{s_{\text{KE}_{4}}^{2}}+\frac{9\sin^{2}\al}{5+\cos 2\al}\eta^{2}\Big)\,.
\end{equation}
with $\e^{2f(\al)}=L^{2}\e^{\phi_{0}/2}\sqrt{5+\cos 2\al}$.
We can write this is in a more compact manner, as follows
\begin{gather}\label{10d metric warped}
	\begin{aligned}
		\dd{s}^{2}&=\dd{s'}^{2}+\dd{\tilde{s}}^{2}_{X_{6}}\\
		&=\e^{2f(x^{u})} g_{\al\beta}(x^{\gamma})\dd{x}^{\al}\dd{x}^{\beta}+g_{st}(x^{u})\dd{x}^{s}\dd{x}^{t}\,.
	\end{aligned}
\end{gather}
where we split the ten-dimensional target space indices $\mu,\nu$ into indices from the beginning of the Greek alphabet $\al,\beta=1,2,3,4$ for the AdS$_{4}$ part and indices from the end of the Latin alphabet $s,t,u=5,\dots,10$ for the $X_{6}$ part. It is then easy to show that the non-zero components of the Riemann tensor are
\begin{gather}\label{Riemann}
	\begin{aligned}
		\tensor{R}{_{\al\beta\gamma}^{\delta}}&=\tensor{R}{^{'}_{\al\beta\gamma}^{\delta}}+g^{st} \partial_{s}f\partial_{t}f\e^{2f}\qty(g_{\gamma\beta}\de^{\delta}_{\al}-g_{\gamma\al}\de^{\delta}_{\beta})\,,\\ 
		\tensor{R}{_{s\al\beta}^{t}}&=\e^{2f}g_{\al\beta}g^{tu}\qty(\tilde\nabla_{s}\partial_{u}f+\partial_{s}f\partial_{u}f)\,,\\ 	
		\tensor{R}{_{s\al t}^{\beta}}&=-\de_{\al}^{\beta}\qty(\tilde\nabla_{s}\partial_{t}f+\partial_{s}f\partial_{t}f)\,, \\
		\tensor{R}{_{\al s\beta}^{t}}&=-\e^{2f}g_{\al\beta}g^{tu}\qty(\tilde\nabla_{s}\partial_{u}f+\partial_{s}f\partial_{u}f)\\
		\tensor{R}{_{stu}^{v}}&=\tensor{\tilde{R}}{_{stu}^{v}}\,,
	\end{aligned}
\end{gather}
with $\tilde\nabla_{s}\partial_{t}f=\partial_{s}\partial_{t}f-\tilde\Gamma^{u}_{st}\partial_{u}f$, and the prime denotes objects evaluated in the AdS$_{4}$ manifold, while the tilde denotes the objects of $X_{6}$.
Now, since the metric is written as a warped product, we can write the mass matrix in the following block-diagonal form
\begin{gather}
	\mathcal{M}_{ab}=\begin{bmatrix}
		\mathcal{M}^{\text{AdS}} & 0 \\
		0 & \mathcal{M}^{X_{6}}
	\end{bmatrix}\,,
\end{gather}
where $a,b=1,2$ are the tangent space indices assigned to the remaining transverse AdS directions and $a,b=5,\dots,10$ are the ones related to the internal manifold $X_{6}$. 
Now, from \eqref{Riemann} it is easy to see that the AdS part of the mass matrix is completely independent of the six-dimensional internal manifold, as it is given by $\mathcal{M}^{\text{AdS}}_{ab}=\tfrac{2}{\ell^{2}}\delta_{ab}$ and what remains is to find an expression for $\mathcal{M}^{X_6}$. Using \eqref{Riemann} we have 
\begin{equation}\label{X6 mass matrix}
	\mathcal{M}^{X_6}_{ab}	=2\tensor{E}{^{s}_{a}}\tensor{E}{^{t}_{b}}\qty(\partial_{s}\partial_{t}f(\al)-\delta_{u}^{5}\Gamma^{u}_{st}\dot{f}(\al)+\partial_{s}f(\al)\partial_{t}f(\al))\,,
\end{equation}
where we denote the derivative w.r.t. the angle $\alpha$ with a dot and we highlight that the Kronecker delta fixes the $u$ index, which in the language of the coordinates on $X_{6}$ refers to the $x^{5}=\alpha$ coordinate. We continue using this notation in the remainder of this section. 
From here it is obvious that the mass matrix component along the $\alpha$-direction will be independent of the SE manifold. Therefore it is more interesting to look at the mass-matrix components with $\{s,t\}\neq5$. Hence, from here on it is suggestive to denote the indices related to the part of the metric that is topologically a warped SE$_{5}$ as $\bar{s},\bar{t}$, i.e.
\begin{gather}
	\dd{s}^{2}_{X_{6}}=G_{st}\dd{x^{s}}\dd{x^{t}}=\frac{3}{2}\e^{2f(\al)}\dd{\al}^{2}+G_{\bar{s}\bar{t}}\dd{x^{\bar{s}}}\dd{x^{\bar{t}}}\,,\qquad \bar{s},\bar{t}=6,\dots,10\,.
\end{gather}
Using this notation, the only possible non-trivial contribution attributed to the Christoffel symbols appearing in \eqref{X6 mass matrix} is coming from  
\begin{equation}\label{Chris tilde}
	\Gamma^{5}_{55}=\dot f(\al)\,,\qquad 	\Gamma^{5}_{\bar s\bar t}=-\frac{1}{3}\e^{-2f(\al)}\dot G_{\bar s \bar t}\,,
\end{equation}
Moreover, since we only have derivatives w.r.t. $\al$ appearing in the mass matrix we may write the components of the metric as
\begin{equation}
	G_{\bar{s}\bar{t}}=G_{\bar{s}\bar{t}}^{(\mathfrak{c})}h^{(\mathfrak{c})}(\al)\,,
\end{equation}
where $G_{\bar{s}\bar{t}}^{(\mathfrak{c})}=G_{\bar{s}\bar{t}}^{(\mathfrak{c})}(x^{\bar s})$ does not depend on the $\alpha$ angle and the $\mathfrak{c}$ in $h^{(\mathfrak{c})}(\al)$ is just a symbol 
that depending on the values of the indices $\bar{s},\bar{t}$, refers to the following functions
\begin{equation}\label{h funct}
	h^{(\text{KE})}(\al)=6\sin^{2}\al\frac{\sqrt{5+\cos 2\al}}{3+\cos 2\al}\,,\qquad	h^{(\eta)}(\al)=\frac{9\sin^{2}\al}{\sqrt{5+\cos 2\al}}\,,
\end{equation}
extracted from \eqref{general warped metric}. Then \eqref{Chris tilde} can be written as
\begin{gather}
	\Gamma^{5}_{\bar s\bar t}=-\frac{\e^{-2f(\al)}}{3}G_{\bar{s}\bar{t}}^{(\mathfrak{c})}\dot{h}^{(\mathfrak{c})}(\al)\,,
\end{gather}
and thus we arrive at the following expression for the mass matrix 
\begin{gather}\label{mass matrix bos}
	\mathcal{M}^{X_6}_{ab}=\frac{2\e^{-2f(\al)}}{3}\qty(2\delta_{5 a}\delta_{ 5 b}\ddot{f}(\al)+\frac{\de_{a\hat{s}}\de_{b\hat{t}}}{h^{(\mathfrak{c})}(\al)}\dot{h}^{(\mathfrak{c})}(\al)\dot{f}(\al))\bigg\lvert_{\alpha=0}\,,
\end{gather}
where the hatted indices $\hat{s},\hat{t}=6,\dots,10$ denote the respective tangent space indices of the space that is topologically a SE$_{5}$ manifold and we recall that the Latin indices $a,b$ denote the flat transverse directions to the worldsheet.\footnote{To improve readability, we removed the bar from the SE$_{5}$ indices.} Note that   $\mathfrak{c}=\eta$ for $\hat{s},\hat{t}=10$ and $\mathfrak{c}=\rm{KE}$ otherwise. Equation \eqref{mass matrix bos} together with $\mathcal{M}^{\text{AdS}}_{ab}=\tfrac{2}{\ell^{2}}\delta_{ab}$ implies that the bosonic mass matrix is indeed diagonal and that it is completely independent of the choice for the KE metric and its contact one-form $\eta$. Hence, the mass spectrum of both the bosonic and fermionic quadratic operators is universal.

%----------------------------------------------------------------------------------------
%	REFERENCES
%---------------------------------------------------------------------------------------- 
\newpage
\bibliographystyle{JHEP}
\bibliography{refs}
\end{document}